\documentclass[aip,notitlepage]{revtex4-1}

\setcitestyle{super}

\usepackage{times}
\usepackage{comment}
\usepackage{graphicx,subfigure}
\usepackage{bm}
\usepackage{amsmath,amsfonts,amsthm,amssymb}
\usepackage{mathtools}
\usepackage[abs]{overpic}
\usepackage{color}
\usepackage[usenames,dvipsnames]{xcolor}
\usepackage{soul}
\usepackage{comment}
\usepackage[breaklinks]{hyperref}
\usepackage{epstopdf, epsfig}
\usepackage{grfext,grffile}
\usepackage[normalem]{ulem}

\newcommand{\bk}{{\bf k}}

\newcommand{\eg}{\emph{e.g.}}

\newcommand{\be}{\begin{equation}}
\newcommand{\ee}{\end{equation}}
\newcommand{\bi}{\begin{itemize}}
\newcommand{\ei}{\end{itemize}}

\newcommand{\TODO}[1]{\textcolor{red}{TO DO: #1}}

\newcommand{\corr}{\color{black}}
\newcommand{\rroc}{\color{black}}

\begin{document}
\graphicspath{{Figs/v1/}}

\title{Inverse cascade anomalies in  fourth-order Leith models}
\author{Simon Thalabard}\email{simon.thalabard@inphyni.cnrs.fr}\affiliation{Instituto Nacional de Matem\'atica Pura e Aplicada -- IMPA, 22460-320 Rio de Janeiro, Brazil}
\affiliation{Institut de Physique de Nice, Universite C\^ote D'Azur et CNRS, Nice 06108, France}
\author{Sergey Medvedev}\email{serbormed@gmail.com}\affiliation{Federal Research Center for Information and Computational Technologies, Novosibirsk 630090, Russia}\affiliation{Novosibirsk State University, Novosibirsk 630090, Russia}
\author{Vladimir Grebenev}\email{vngrebenev@gmail.com}\affiliation{Federal Research Center for Information and Computational Technologies, Novosibirsk 630090, Russia}
\author{Sergey Nazarenko}\email{Sergey.NAZARENKO@unice.fr}\affiliation{Institut de Physique de Nice, Universite C\^ote D'Azur et CNRS, Nice 06108, France}

\begin{abstract}
We analyze a family of fourth-order non-linear diffusion models corresponding to local approximations of 
4-wave kinetic equations of weak wave turbulence. We focus on a class of parameters for which a dual cascade behavior is expected with an infrared finite-time singularity associated to inverse transfer of waveaction. This case is relevant for wave turbulence arising in the Nonlinear Schr\"odinger model and for the gravitational waves in the Einstein's vacuum field model.
We show that inverse transfer is not described by a scaling of the constant-flux solution but has an anomalous scaling.
We  compute the anomalous exponents and analyze their origin using the theory of  dynamical systems.
\end{abstract}

\maketitle

\section{From kinetic equations to non-linear diffusions}
\paragraph*{Single cascade and second-order diffusions.}
The  differential approximation model (DAM) introduced  by Cecil Leith  in the context of two-point spectral closures for  fully-developed homogeneous isotropic turbulence  \cite{leith1967diffusion} provides a class of  tractable models
with two desirable features: \emph{(i)} compatibility with thermodynamics, that allows both for the Lee equilibrium states \cite{lee1952some}and the Kolmogorov cascade solutions to emerge and \emph{(ii)} connection to kinetic equations, \eg\, 
spectral closures featuring integro-differential approximations  under a Markovian approximation are known to reduce to  non-linear diffusions when interactions are restricted onto well-chosen subclasses of local triads \cite{leith1967diffusion,rubinstein2017leith,clark2009reassessment}.
 Leith's approximation  can be thought of as a type of non-linear Fourier law, where the energy flux is written as a non-linear function of the energy spectrum and of its spectral derivative. As such, the approximation is of \emph{second-order} and it
 provides a flexible framework to address fluid systems whose physics can be reduced to   single conservation laws.
While Leith's original model is too crude to produce quantitatively precise results for strong turbulence\cite{orszag1973test}, 
it describes robust qualitative features of turbulence 
beyond the properties that are hardwired into it 
, \emph{e.g.} the Kolmogorov and the thermodynamic steady states. In particular,  the model has a stationary solution in the form of a mixed Kolmogorov-thermodynamic  state --the so-called ``warm cascade"~\cite{connaughton2004warm}.
Further, its generalization  has proven  fruitful to get qualitative insights in various turbulent settings. For atmospheric turbulence, it was used to substantiate the origin of the Nastrom-Gage spectrum of atmospheric turbulence  in terms of a dual cascade state, \corr involving energy and enstrophy  flowing through a common  inertial range of scales~\cite{Lilly,Morel}\rroc. 
\corr Another salient  generalization of Leith model coupling    superfluid and  normal-fluid components  via a mutual friction has allowed to unveil interesting new scaling states  relevant for superfluid turbulence ~\cite{LvovVolovik,LvovSkrbek}. Besides, 
Leith type models also found relevance in
the context of passive transport and wave turbulence \cite{thalabard2015anomalous}, where they have proven  useful to address non-stationary dynamics. \rroc

Among salient features, second-order models with finite capacity  are now known to feature self-similar finite-time blow-up with anomalous scaling; this observation holds true for a wide class of systems \cite{thalabard2015anomalous} including the original Leith model \cite{connaughton2004warm}, but also  systems  featuring inverse cascade behavior \cite{galtier2019nonlinear}. What makes the diffusion approximation particularly valuable is the fact that unlike other classes of closures \cite{l1998improved,campolina2018chaotic}, the diffusion structure allows for the mathematical analysis of the anomalous transients. In particular, it was recently found that those  can be analyzed in terms of self-similar solutions of the second-kind \cite{GNMCh2014}, where a self-similar profile develops in finite time over an infinite  range of scales, featuring a sharp  front on one side and an anomalous algebraic decay on the other side of the spectral range. 

\paragraph*{Dual cascades and  fourth-order diffusions.}
Generalized DAM describing single-conservation laws can be heuristically derived from dimensional analysis \cite{nazarenko2011wave} to gain insights on the behavior of the relevant kinetic equations. An obvious shortcoming of such second-order DAM is their inability to address dual cascade scenario---a building block of wave-turbulence. To model systems with two  conservation laws within a diffusion approximation,  one has to rely on another class of DAM, later referred to as  \emph{fourth-order models}.
Beyond the  case of two-dimensional (2D) turbulence~\cite{l2006differential},  fourth-order DAM find  applications in wave systems relevant for classical fluids, e.g.  gravity waves on waver surface~\cite{Hasselmann}. Non-classical examples include Kelvin waves on quantized vortex lines \cite{DAMkelvin}, gravitational waves (GW)~\cite{galtier2017turbulence}, waves in fuzzy dark matter\cite{skipp2020wave} or Bose-Einstein condensates (BEC)~\cite{connaughton2004kinetic}. 

Recently, it was suggested that self-similar solutions of the second kind could also prove relevant to describe blow-up condensation scenarios in a class of fourth-order diffusion \cite{connaughton2004kinetic}, which include as physical examples both the GW and wave turbulence in BEC described by the Nonlinear Schr\"odinger (NLS) equation. In the case of GW,  the relevance of self-similar solutions and anomalous scaling  was suggested \cite{galtier2019nonlinear} from the analysis of a companion second-order DAM following the general approach described in Refs.~\cite{thalabard2015anomalous,nazarenko2016self}.  Construction of relevant second-order DAM is however not necessarily unambiguous: In particular, the reduction strategy from fourth to second order is likely to produce nonphysical results, a feature previously observed in the context of 2D
turbulence\cite{leith1968diffusion,l2006differential}.
Beyond numerical observations and analogies, it remains therefore unclear whether a proper dynamical system analysis can be extended to fourth-order systems, to provide precise characterization of finite-time blow-ups. In the present paper, we  study such a finite-time blow-up  using  numerical simulations of the fourth-order DAM, and characterize its properties in terms of a  self-similar Ansatz; \corr This allows to reduce the problem to the analysis of  a four-dimensional (4D) dynamical system \rroc.

The organisation of the paper is as follows. In Section~\ref{S2},
we present a one-parameter family  of DAM,  which provide  fourth-order approximations of the wave kinetic equation for four-wave $(2 \to 2)$ interactions, and prove  in particular relevant for GW and waves in the NLS model. 
 Section~\ref{S3} focuses  on the three-dimensional (3D) GW and the 3D  NLS cases; We describe  numerical simulations of the evolution problem with compact supported initial spectra,  revealing  finite time blow-ups at the infrared front, characterized by  anomalous spectral exponents. 
 In Section~\ref{S4}, we discuss the anomalous scaling in terms of simplified second-order DAM and self-similar solutions of the second-kind. In Section~\ref{S5},
 we extend the approach to the fourth-order case. 
\corr We characterize the anomalous exponent in terms 
 of  a nonlinear eigenvalue problem, and  formulate the problem in terms of bifurcation analysis within an autonomous 4D dynamical system. \rroc
 In Section~\ref{S6}, we employ
 numerical  continuation algorithms to explicitly determine the  anomalous exponent, in particular for, but not limited to, the  3D GW and the 3D NLS models.
Section~\ref{sec:conclusion} exposes concluding remarks.

\section{The fourth-order model}\label{S2}
\subsection{The model and notations.}

In the present paper, we will study the fourth-order DAM that describes the  conservation of the waveaction density (spectrum) $N(\omega,t)$ in frequency space as 
\be
	\begin{split}
	\partial_t N+\partial_{\omega} Q=0 , 
	\;\text{with }\;	Q (\omega,t) :=-\partial_{\omega} K,\;\; \text{and}\;
	K(\omega,t) :=\omega^{5-r}N^4  \partial_{\omega\omega}^2 \left(\omega^{r}N^{-1}\right).
	\end{split}
	\label{eq:LeithFour}
\ee
The coefficient $r=d/\alpha-1$ is a physical parameter involving  the spatial dimension $d$  and  the exponent $\alpha$ associated to  the dispersion relation of the interacting waves (see next paragraph for examples).
The system 	(\ref{eq:LeithFour}) is a \emph{fourth-order approximation}, as  the flux is  written as a non-linear Fourier law involving up to the third-derivative of the waveaction spectral density $N$. The resulting dynamics involves terms  up to the fourth derivative of $N$, hence the name.
In previous works, the same equation is sometimes found  written in terms of the $d$-dimensional $\bk$-space waveaction spectrum $n(\bk,t)$.
The $\omega$-space density is used when the spectrum is isotropic, i.e. independent of the direction of $\bk$; it 
relates to the $d$-dimensional spectrum through 
integration over the angular directions in $\bk$-space and
multiplying by the Jacobian $|\partial k/\partial \omega|$, 
where the function $\omega(\bk,\alpha) \propto k^\alpha$ (with $k=|\bk|$) specifies the dispersion relation for the wave frequency in terms of the wave vector length and  coefficient $\alpha$. This gives
\begin{equation}
n(\bk,t)\propto \omega^{1-d/\alpha}N(\omega).
\end{equation}

\subsection{Interpretation: local approximation to kinetic equations.}
	
System~(\ref{eq:LeithFour}) describes a one-parameter family of DAM depending upon the the coefficient $r := d/\alpha-1$.
The model connects to the 
four-wave kinetic equation
	\be
    \label{eq:kinetic}
		\begin{split}
		\partial_t n(\bk,t)=  \int_{{\mathbb R^{3d}}} &\text{\bf d} \bk_*\text{\bf d}\bk'\text{\bf d}\bk'_* \, |T_{\bk\bk_* \to \bk'\bk'_*}|^2 n n' n_* n_*' \left( \dfrac{1}{n}+\dfrac{1}{n_*}-\dfrac{1}{n'}-\dfrac{1}{n_*'}\right)\\
				&\delta(\omega+\omega_*-\omega'-\omega'_*) \,\delta(\bk+\bk_*-\bk'-\bk'_*)
		\end{split}
	\ee
	{with} frequency $\omega= k^\alpha$ {and} interaction coefficient 
$		 	
T_{\bk \bk_* \to \bk' \bk'_*} \equiv T({\bk, \bk_*, \bk', \bk'_*})$ is a
homogeneous function of degree $\gamma$: $T({\lambda\bk, \lambda\bk_*, \lambda\bk', \lambda\bk'_*}) =\lambda^\gamma T({\bk, \bk_*, \bk', \bk'_*})
$ for any $\lambda>0$.
 Assuming statistical isotropy,  $n({\bf k}) \equiv n({ k})$, as well as a strong localisation of   $T_{\bk\bk_* \to \bk'\bk'_*}$  on neighboring wavenumbers with $\bk \approx \bk_* \approx \bk' \approx \bk'_*$,  the kinetic equation~(\ref{eq:kinetic}) becomes  the slightly generalized two-parameter version of  Eq.~(\ref{eq:LeithFour}) \cite{nazarenko2011wave}, 
\be
	\begin{split}
	\partial_t N+\partial_{\omega} Q=0 , 
	\;\text{with }\;	Q (\omega,t) :=-\partial_{\omega} K,\;\; \text{and}\;
	K(\omega,t) :=\omega^{5-r+2g}N^4  \partial_{\omega\omega}^2 \left(\omega^{r}N^{-1}\right),
	\end{split}
	\label{eq:LeithFour_gen}
\ee
where $g=\gamma/\alpha$.
Then, Eq.~(\ref{eq:LeithFour}) is  obtained by restricting to the  case $g=\gamma=0$: it is in particular relevant for the GW turbulence  in the Einstein vacuum model, as well as the NLS wave turbulence, with application both in nonlinear optics and BEC. Below, we will focus on the systems with $\gamma=g=0$, but for completeness  we present a list of examples including  finite $g$ cases in  Table \ref{tab:examples}.
\renewcommand{\arraystretch}{1.5}
\begin{table}[h!]
\setlength{\tabcolsep}{18pt}
    \centering
    \begin{tabular}{cccccc}
         System &  $d$ & $\alpha$ & $\gamma$ & $r$ & $g$\\
         \hline
         \hline
         Deep water surface gravity waves \cite{nazarenko2011wave} &  2  & $\frac{1}{2}$ & 3 &  3  & 6\\    
        \hline
         Self-gravitating dark matter\cite{skipp2020wave} &  2 \; or \;3 & 2 & -2 &  0 \;or \; $\frac{1}{2}$ & -1\\
         \hline
         Bose-Einstein condensates (NLS) \cite{nazarenko2011wave} &  2 \;or \; 3 & 2& 0&  0 \;or \; $\frac{ 1}{ 2}$ & 0\\
         \hline
         Gravitational waves in vacuum \cite{galtier2017turbulence,galtier2019nonlinear} &  2 \;or \; 3 & 1& 0&  1 \;or \; 2  & 0\\
        \hline
        \hline
        \end{tabular}
    \caption{List of relevant parameters in examples of four-wave  turbulent systems.
    }
    \label{tab:examples}
\end{table}

\paragraph{Kraichnan-Lee vs Kolmogorov-Zakharov solutions.}
System (\ref{eq:LeithFour_gen}), which includes System (\ref{eq:LeithFour}) as a special case,  is explicitly written in the form of a conservation law for the waveaction $ N$, 
but it also implies conservation of the energy $E:=\omega N$. Indeed, straightforward algebraic manipulations  lead to  
\be
	\partial_t E + \partial_\omega P = 0  \text{  with  } P := K+\omega Q.
	\label{eq:LeithFour_E}
\ee
This formulation  is naturally equivalent	to the formulation of Eq.(\ref{eq:LeithFour_gen}). Yet, one interesting feature of the fourth-order DAM (\ref{eq:LeithFour_gen},\ref{eq:LeithFour_E}) is the fact that they truly mimic the physics of dual-cascade systems. At an elementary level, this can be seen from the observation that  they feature four different types of pure scaling steady solutions 
$N \propto \omega^{-x}$. The first types are the   equilibrium states with  $P=Q=0$ leading to
\be
	  x=x_N=-r \;\text{ (Equipartition of $N$)}, \hspace{1cm}\text{or }\hspace{1cm} x=x_E=1-r  \;\text{(Equipartition of $E$)},
\ee
respectively. These solutions
 are asymptotics of  a more general Kraichnan-Lee (KL) equilibrium solution $N = \omega^r/(A \omega +B)$ with $A,B=$~const, also known as Rayleigh-Jeans in the wave turbulence literature \cite{kraichnan1967inertial,nazarenko2011wave}.
The other types  are the Kolmogorov-Zakharov (KZ) stationary solutions corresponding to the constant fluxes of energy and  waveaction respectively: 
\be
\label{kzPQ}
N= C_P P^{1/3} \omega^{-x_P} \hspace{1cm}\text{and }\hspace{1cm} N= C_Q (-Q)^{1/3} \omega^{-x_Q} 
\ee
with exponents
\be
	x=x_P=1 +2g/3 \;\text{ ( $Q=0,P \neq 0$)   } \hspace{1cm}\text{and }\hspace{1cm}  x=x_Q=\frac23(1+g) \text{ \;( $Q\neq 0, P = 0$)}
\ee
and dimensionless (KZ) constants $C_P $ and $C_Q $.

The pure direct cascade and the pure inverse cascade power law spectra can in principle be realized only if they correspond to the flux directions $Q<0$ and $P>0$ compatible with the  standard arguments of the Fjortoft type \cite{nazarenko2011wave}. This condition is satisfied if and only if
$2-3d/2\alpha <g$ which is true for the GW, the 3D NLS and the deep water gravity waves, but fails for the the waves in self-gravitating dark matter and 2D NLS.  When this condition fails, the pure cascade states cannot be realised and mixed cascade-thermodynamic (``warm cascade") states are expected.


Even though the KZ scalings were ``built into"  the fourth order DAM, the latter have predictive power beyond these scalings and, for instance, give predictions for the constant pre-factors of the KZ spectra~\cite{l2006differential}. 
Indeed,
substituting the two spectra \eqref{kzPQ} into the expressions of the respective fluxes and taking the ratio of the resulting equations, we get
\be
\label{cpcq}
\frac{C_P^3}{C_Q^3} = \frac{(r+2/3+2g/3)(r-1/3+2g/3)}{(r+1+2g/3)(r+2g/3)}.
\ee
Note that, to be fully predictive, the right-hand side of the 
model (\ref{eq:LeithFour_gen}) should contain an order-one pre-factor which depends on a specific wave system. Such a pre-factor however drops out from the ratio  $C_P/C_Q$, and  in particular, we get $C_P/C_Q=0.638$ for the 3D NLS,  $C_P/C_Q=0.905$ for the 3D GW and  $C_P/C_Q=0.970$ for the deep water gravity wave turbulence.\\

\corr
\section{ Numerical simulations of the fourth-order DAM}
\rroc
\label{S3}
In the present work, we restrict our attention to freely evolving wave turbulence (without forcing or/and dissipation).  Here, we use the  numerical simulations  of the fourth-order DAM (\ref{eq:LeithFour})
 to characterize the free evolution of an initial condition that features initial  compact support around a frequency $\omega_i$ in frequency space.
Such initial condition has finite energy and  waveaction, $\int \omega N d\omega, \int N d\omega <\infty$, and as such, the system could, in principle, propagate the waveaction towards $\omega\to 0$ through a constant-flux solution, as the corresponding scaling $N \sim \omega^{-2/3}$ entails finite-capacity on the infra-red end, that is $\int_{0}^{\omega_i} N < \infty$. The compact initial condition cannot,  however,  sustain a direct cascade towards $\omega \to \infty$:  the latter indeed entails the scaling $N \sim \omega^{-1}$ on the ultra-violet side $\omega \to \infty$, and this scaling requires an infinite  physical space density of waveaction for the spectrum   to extend up to $\omega\to\infty$.\\

\begin{figure}
	\includegraphics[width=\columnwidth]{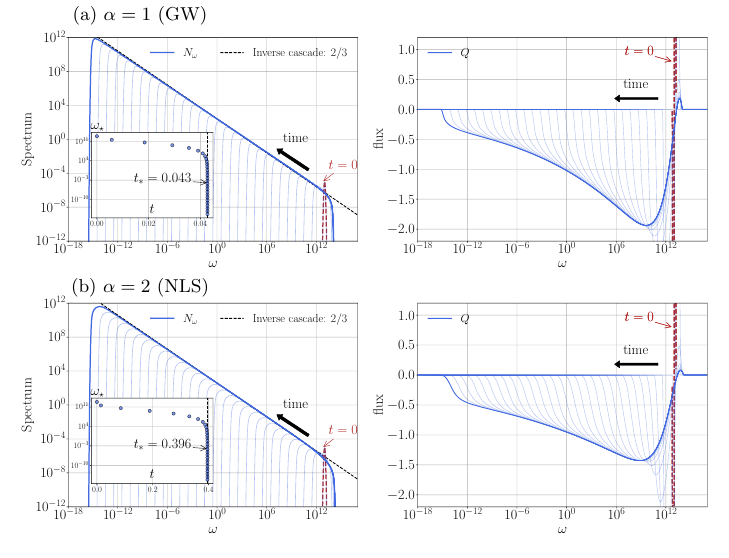}
	\caption{{\bf Numerical simulations  for the fourth-order DAM: spectra and fluxes.} Spectra (left) and fluxes (right) obtained from direct numerical simulations of \corr (a) the 3D GW, and (b) the 3D NLS,  starting from the narrow Gaussian profile about $\omega=10^{12}$ displayed in red\rroc. The insets  in the left panels show the position of the left front against time, revealing finite-time blow-up. \corr See also videos online\rroc.}
\label{fig:DNS}
\end{figure}

\corr
To perform numerics of DAM, it is important to take extra care of the tendency to form sharp propagating fronts with discontinuous derivatives which typically lead to numerical blow-up when a simple differentiation scheme is used. To tackle this obstacle, we rely on smooth noise-robust differentiators, and use a log-discretization in the frequency space to allow for extended computational range. Further technicalities related to our numerical methods  are described in Appendix \ref{sec:dnsdetails}.
The numerical results shown in Fig. \ref{fig:DNS}  \rroc reveal that 
 the waveaction apparently cascades towards the infra-red  end and exhibits scaling close to 2/3. \corr This corresponds to constant flux solutions, and the direction of the cascade is indeed compatible with the heuristics of Section \ref{S2}. One also observe that the waveaction  reaches $\omega \to 0$ in  finite time $t_*$. \rroc For the same initial condition, the blow-up time $t_*$ is ten times smaller in the 3D GW case than in the 3D NLS case, as   shown in the insets within the left panels of Fig. \ref{fig:DNS}.

\begin{figure}
	\includegraphics[width=\columnwidth]{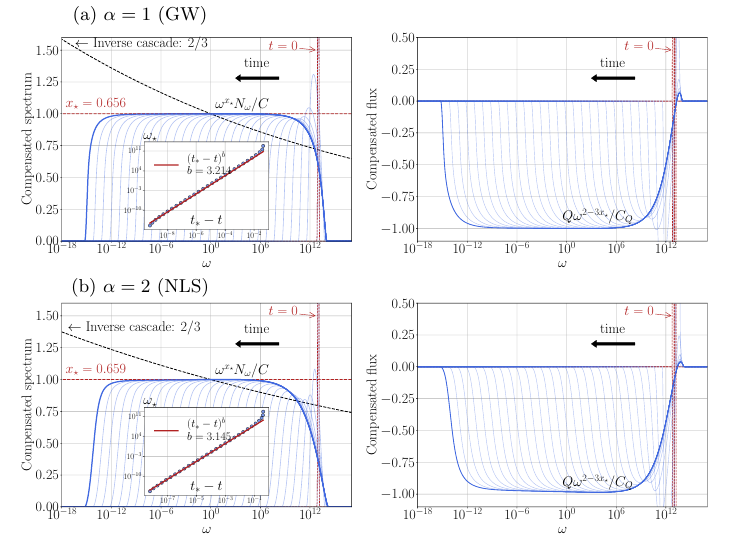}
	\caption{{\bf Numerical simulations  for the fourth-order DAM: compensated spectra and fluxes.} Same as Fig. \ref{fig:DNS}, but this time the spectra and the fluxes are suitably compensated  with carefully chosen exponents $x_*\simeq 0.656$ for GW and $x_*\simeq 0.659$ for NLS, \corr and shown in log-lin coordinates.\rroc The insets  in the left panels show the position of the left front against time $t_*-t$ in log-log coordinates. \corr See also videos online \rroc.}
	\label{fig:DNS_compensated}
\end{figure}

Closer inspection however reveals that those finite-time transients are anomalous: The fluxes  converge towards a profile which is not a constant but rather an increasing function of $\omega$, as shown in the right panels of  Fig.\ref{fig:DNS}.  The compensated spectra  shown in Fig. \ref{fig:DNS_compensated} reveal small  deviations from pure scaling solutions. Instead of  the 2/3 KZ scaling, the  numerics feature the behavior $N\sim \omega^{-x_*}$, with $x_*\simeq 0.656$ for the GW and $x_*\simeq 0.659$ for the 3D NLS. In both cases, the deviations to pure KZ scaling is small (1.6\% and 1.2\% respectively), but measurable. 
Besides, the infra-red blow-up is algebraic. The front reaches $\omega=0$ following the apparent power-laws $\propto (t_*-t)^b$ with $b\simeq 3.145$ in the 3D GW case and $b\simeq 3.214$ in the 3D NLS case.
The remainder of the  paper aims at characterizing such anomalous scalings.

Out of the four-wave systems listed in the table, we choose to study only the GW, the 3D NLS and the systems with $g=0$, $d=3$ and $1/2<\alpha<9/4$, in particular including the gradual transition from the GW ($\alpha=1$) to the 3D NLS ($\alpha=2$) systems. We do not here study the deep water gravitational waves as this system does not have finite capacity at the infra-red end, and therefore does not exhibit an anomalous inverse cascade scaling. Also, we do not study the waves in the self-gravitating dark matter because, as we argued earlier, it has an ordering of the exponents of the stationary power-law solutions inconsistent with the Fjortoft dual cascade argument, meaning that they correspond to a different class in which ``warm cascades" are expected. 

\vspace{.5cm}

\section{Deficiency of the reduction to second-order DAM}\label{S4}
\label{sec:Failure}
The anomalous scalings observed in Fig.~\ref{fig:DNS_compensated}
are reminiscent of  behaviors  previously observed in second-order DAM \cite{connaughton2004warm, thalabard2015anomalous}.   In particular,  existence of anomalous transients for  the 4th-order DAM for the  GW turbulence combined with algebraic finite-time blow-up was suggested from the theoretical and numerical analysis of a companion second-order DAM\cite{thalabard2015anomalous,nazarenko2019focusing}. 
  Such reduction to second-order DAM can be derived heuristically by engineering the waveaction flux to \emph{(i)} feature one derivative only, \emph{(ii)}  feature  homogeneity  $\propto N^3$ as the fourth-order DAM,  \emph{(iii)}  yield constant-flux scaling $\propto \omega^{-2/3}$  and \corr \emph{(iv)} yield equilibrium scaling \rroc matching either equipartition of $E$ or $N$ or energy flux.
This scheme prescribes the one-parameter family of  second-order DAM
 \be
\partial_t N + \partial_\omega Q=0  \text{  with  } Q:= -\omega^{3+\rho }N^2 \partial_{\omega} N \omega^{-\rho}.
\ee
By construction,  the model has constant flux solution $N\sim \omega^{-2/3}$, and  equilibrium solutions $N\sim \omega^{\rho}$. 
Constant $\rho$ represents an effective dimension, whose value  determines the complimentary equilibrium scaling. The three relevant choices for $\rho$  are $\rho=r$ (waveaction equipartition),  $\rho=r-1$ (energy equipartition)  or $\rho=-1$ (constant energy flux).
The crucial observation is that  among those three choices, only the third choice $\rho=-1$ is compatible with inverse cascade of waveaction
$N= C\omega^{-2/3}$ with $C>0$. This comes from the fact that  the associated flux is  $Q=C^3  (\rho+2/3)$ --   negative only provided $\rho<-2/3$.
For both the GW and the 3D NLS, the solutions $\rho=r$ and $\rho =r-1$ do not fulfill this condition. This means  that both the GW and the 3D NLS reduce to the same second-order DAM:
 \be
	\partial_t N + \partial_\omega Q =0,  \text{  with  } Q:= -\omega^{2 }N^2 \partial_{\omega}( N \omega).
	\label{eq:2order_GWNLS}
\ee
As such, the reduction from fourth to second-order model simplifies the system but   cannot explain the weak but measurable difference between
the anomalous scalings observed  in the GW and the 3D NLS systems.
Indeed, the investigation of Eq.(\ref{eq:2order_GWNLS})  reveals existence of  anomalous transients  $N\sim \omega^{-x_*}$;  the anomalous scaling is   $N\sim \omega^{- 0.65169}$, only about $2\%$ shallower than KZ, but clearly distinguishable from it in numerics \cite{galtier2019nonlinear}. \corr While the anomalous scaling qualitatively matches the observations for the fourth-order DAM, it  does not quantitatively correspond to either of the scalings observed  in Fig.\ref{fig:DNS_compensated}. We note that  the anomalous scaling in the second-order DAM was previously elucidated \cite{galtier2019nonlinear}, and we  recall its origin in  Appendix \ref{sec:GW_2}. \rroc

Another obvious deficiency of the second-order model is seen at the level of the stationary KZ solutions: One can easily  check that the second-order DAM predicts equal values of the KZ constants, \emph{e.g.} $C_P=C_Q$. \corr This is unlike the more realistic fourth-order DAM, which  predicts different values of the KZ constants($C_P=0.638 C_Q$ for the 3D NLS and $C_P=0.905 C_Q$ for the 3D GW), as seen from Eq.~\eqref{cpcq}.
\corr
Those undesired deficiencies of the second-order model motivate the detailed investigation of the fourth-order DAM.
\rroc
\section{Anomalous transients as a self-similarity of the second-kind}
\label{S5}

\subsection{The nonlinear eigenvalue problem.}
We now look for self-similar solutions of the second-kind, which feature a self-similar profile $F$ invading the full infra-red range in finite time $t<t_*$:
Introducing the propagating front  $\omega_*$ and the self-similar variable $\eta:= \omega/\omega_*$, those solutions 
take the form
\be
	\begin{split}
				&N(\omega,t)= \omega_*^{a} F(\eta)  \;\; \text{with } \;\omega_* \propto (t_*-t)^b,\quad b>0\\
		\text{ and the boundary conditions}&\text{ $F(\eta) \underset{{\color{black}\eta\to} 1}\to 0$ ``sufficiently smoothly'' and $F(\eta) \underset{{\color{black}\eta\to}\infty}{\sim} \eta^{-x}$}.
	\end{split}
	\label{eq:selfsimilar2}
\ee
Here, ``sufficiently smoothly'' means that we select a physically relevant solution such that at the front, where the spectrum vanishes, the fluxes of the waveaction and the energy also vanish (see more about this condition below). The power law asymptotics at the ultra-violet side is \corr an expected feature \rroc of  self-similar solutions of the second type: finding the power index $x=x_*$ such that the solution satisfies the boundary conditions at both ends constitutes the ``nonlinear eigenvalue problem"  \cite{connaughton2004warm,GNMCh2014,thalabard2015anomalous,bell2017self,SEMISALOV2021105903}.

We recall that the coefficient $r$  depends on the physical parameters as  $r:=d/\alpha-1$.
The condition that $F \sim \eta^{-x}$  must be a valid asymptotic solution for large $\eta$ requires $a=-x $.
The finite-time  convergence of the front towards $\omega\to 0$ requires non-negativity of the coefficient $b>0$. 
The specific expression of $b$ is obtained   by inserting the Ansatz (\ref{eq:selfsimilar2}) into the fourth-order system (\ref{eq:LeithFour})
and requiring that the time variable drops out of the resulting equation;
this gives 
\be b=\frac{1}{2x-1} >0.
\ee
and implies in particular $x>1/2$.
The  procedure then also  yields  the fourth-order ordinary differential equation for $F$
\be
		  b(xF + \eta \partial_\eta F)= \partial_{\eta\eta}\left( \eta^{5-r}F^4\partial_{\eta\eta} \left(\eta^r F^{-1}\right)\right).  
		\label{eq:selfsimilarF}
\ee

We define the anomalous value $x_*$ is the exponent $x$ allowing to solve (\ref{eq:selfsimilarF}) with suitable two-end boundary conditions mentioned above. To specify those in more detail, let us slightly  abuse  notations to introduce the self-similar fluxes 
\be
	Q(\eta):= -\partial_\eta K, \;\; P(\eta):=K+\eta Q \;\; \text{where }  K(\eta):= \eta^{5-r}F^4\partial_{\eta\eta} \left(\eta^r F^{-1}\right).
	\label{eq:bc_infty}
\ee
Further, let us denote $G=F'$.
On the ultraviolet end, algebraic decay $F\sim \eta^{-x}$ prescribes 
\be
\label{eq:bc-inf}
	F, G, P, Q \to 0 \; \text{as}\; \eta\to \infty.
\ee
Note also that for $x>1/2$, scaling solutions $F\sim \eta^{-x}$  the ratio of the r.h.s. of (\ref{eq:selfsimilarF}) to each of the terms
on the l.h.s. asymptotically vanishes as $\eta\to \infty$, 
so they are  indeed  valid asymptotic solutions (see Appendix \ref{sec:etainf}).

On the infrared end $\eta \to 1$, 
we seek solution in form $ F  \to  C(\eta-1)^{\mu} $ for some positive constants $C$ and $\mu$. This expression provides an asymptotic solution  of Eq.~(\ref{eq:selfsimilarF}) with vanishing fluxes $P$ and $Q$ if $\mu=3/2$ and $C=\sqrt{{8b}/{75}}$. \corr This can be seen by the direct substitution of such a form into Eq.~(\ref{eq:selfsimilarF}),  
applying the conditions $P=Q=0$ at $\eta=1$, and retaining the leading order in the expansion in small $ (\eta-1) $ (see Appendix \ref{sec:front}).  Thus we have \rroc
\be
	F \to C (\eta-1)^{3/2},\; G \to \frac{3}{2}C (\eta-1)^{1/2},\; P,Q  \to -\frac{75}{8}C^3 (\eta-1)^{3/2} \;\;\text{as}\; \eta\to 1. 
	\label{eq:bc_one_firstorder}
\ee
hence implying the sharp front behavior
\be
	F, G, P, Q \to 0 \; \text{as}\; \eta\to 1. 
	\label{eq:bc_one}
\ee

\subsection{Reduction to four-dimensional autonomous system.}
To analyze 
solutions of 	Eq.~(\ref{eq:selfsimilarF}),
we now introduce the rescaled variables $f,g,p,q$, defined through 
 		\be
			F=f\eta^{-1/2}, F'=g\eta^{-3/2}, P=p \eta^{3/2}, Q=q\eta^{1/2},
		\ee
 which are now considered as functions of the time-like variable $\tau:=\log \eta$ ranging from $0$ to $\infty$.
In terms of the $f,g,p,q$ variables, Eq.~(\ref{eq:selfsimilarF}) becomes the 4D autonomous dynamical system
 		 \be
 		  \left\{
 			\begin{aligned}
 		 		 f'(\tau) &= g+ \frac{1}{2}f\\
 		 		 f^2 g'(\tau) &= q-p+2fg^2 +\left(\frac{3}{2}-2r\right)gf^2 +r (r-1) f^3 \\
 		 		  p'(\tau) &= -\frac{3}{2}p-b(xf+g)\\
 		 		 q'(\tau) &= -\frac{1}{2}q-b(xf+g)
 			\end{aligned}
 			 \right.,
 			\label{eq:autonomous4}
 		 \ee
with the primes describing derivatives  with respect to the   time-like  variable $\tau$. 
	The boundary conditions  
 	\eqref{eq:bc-inf}--(\ref{eq:bc_one}) now become
 		 \be
			f=g=p=q=0 \;\;\text{at} \;\;\tau =0,\;\;\;\;\text{and}\;\;\;\;
 		 	 f,g,p,q \to 0 \;\;\text{as} \;\;\tau \to \infty.
 		 \ee
 		 
 		 \corr It is easy to see that the  origin $f=g=p=q=0$ is a fixed point of our dynamical system (see the detailed  analysis of the fixed points in Appendix \ref{sec:fixed}). \rroc 
In other words,  solving Eq.~(\ref{eq:selfsimilarF}) with the specified boundary conditions reduces to  searching for a  limit cycle passing through the fixed point at the origin, namely 
a homoclinic cycle. Our conjecture is that such a cycle exists only for a single value of the exponent $x=x_*$: This value
leads to a global homoclinic bifurcation, and finding it is the essence of the nonlinear eigenvalue problem at hand. For $x\ne x_*$ no homoclinic trajectory passing through  the origin exist: The orbits originating at the origin never return to the origin. Since the homoclinic cycles have 
infinite time periods and since they are realised for $x=x^*$, we will later refer to such cycles as $\infty_*$-cycles.\\


\section{Chasing the $\infty_*$-cycles}
\label{S6}
\subsection{The $\infty_*$-cycles in the second-order models.}
The idea of solving the nonlinear eigenvalue problem by connecting its solution to the presence of  a global bifurcation in a suitable associated autonomous dynamical system was previously done for  the second-order DAM \cite{GNMCh2014,thalabard2015anomalous}, \corr including but not limited to to Leith's original model \rroc.
The following scenario was then discovered: When changing the parameter $x$, one first observes change of stability of an isolated fixed point resulting in \corr the creation of a limit cycle through a Hopf bifurcation. As  $x$ is further varied, such a cycle grows until it simultaneously  collides at with two isolated fixed points, one for each boundary conditions. 
In technical words, this scenario is that of a global bifurcation giving birth to a heteroclinic cycle, namely an $\infty_*$ cycle composed of two   heteroclinic orbits. One of the orbits is the solution of the nonlinear eigenvalue problem, and this  determines $x_*$.
Solving the nonlinear eigenvalue problem for the second-order DAM therefore also boils down to chasing an $\infty_*$-cycle;
The property that the global bifurcation creates a heteroclinic cycle rather than  a homoclinic one is not universal, and merely depends on the choice of reduced variables. \footnote{In Appendix~\ref{sec:GW_2}, we show that upon using variables akin to $f,g,p,q$ variables which we introduced to address   the fourth-order DAM, the global bifurcation which determines $ x_*$ for the second-order DAM describes the appearance of a homoclinic cycle.}
 The bifurcation theory for 2D dynamical systems guarantees the existence and the uniqueness of the global bifurcation. While $x_*$ cannot be determined analytically,  the theory provides bounds for its value, which has to be  smaller than the Hopf  value $x_H$ and  greater than the KZ exponent  \cite{GNMCh2014,zamm}. 
 

\subsection{The $\infty_*$-cycles in the fourth-order models.}
\corr
\paragraph{Methods.}
Our aim is to identify the $\infty_*$-cycles for the one-parametric family of fourth-order DAM and associated 4D dynamical system (\ref{eq:autonomous4}), obtained by varying the the coefficient $\alpha \in [0.5,2.25]$, with  the other parameters $g=0, d=3$ prescribed as in the 3D NLS and the 3D GW systems; see Table \ref{tab:examples}.
Bifurcation theory  being less exhaustive for  dynamical systems in 4D than in 2D, we therefore rely on  several additional tools to identify the $\infty_*$-cycles arising in the 4D  system (\ref{eq:autonomous4}): In addition to   the  local analysis described in  Appendix \ref{sec:fixed}, our description uses the classification of co-dimension 1  bifurcations of limit cycles,  as well as numerical  continuation algorithms and the Simpson strategy described in Appendix~\ref{sec:continuation}. 

Unlike  2D, where limit cycles are found to possess stable direction in reverse time, the limit cycles in 4D have unstable directions in both forward and reverse time.  As such, only  a set of initial conditions with zero-measure  results in trajectories attracted to the cycles. This motivates the use of numerical continuation algorithms rather than shooting methods, in order to identify and track the cycles. We specifically rely on the PyCont software from the  PyDSTool library \cite{clewley2007pyds}. 

\paragraph{Generic vs non-generic cases.}
The main outcome of our  analysis  is the conjecture that  similar to the 2D case, there exists  a unique global bifurcation for each member of our  $\alpha$-parametric family of 4D models, each time leading  to the creation of an $\infty_*$-cycle at some prescribed value of $x_*$.  Due do  our specific choice of variables $f,g,p,q$, the $\infty_*$-cycle stems from a homoclinic bifurcation. The specific scenarios leading to such global bifurcation however depend on the 4D system being or not \emph{generic}.
\begin{enumerate}
\item \emph{Generic systems} are obtained  for  $\alpha<2$, or alternatively $r>1/2$, in which case the reduced system (\ref{eq:autonomous4}) possesses two isolated fixed points, namely the origin and a point $P_+$, which undergoes a Hopf bifurcation for some value  $x=x_H$.  This includes the GW system. 
Cycles are found to exist only in between the Hopf value and the KZ value $x_{Q}=2/3$, hereby bounding the value of $x_*$ as in the 2D case.
\item \emph{Non-generic systems} are obtained for  $\alpha \ge 2 $, or alternatively $r \le 1/2$: No isolated fixed point exist except for the  origin. Cycles then exist only for values of $x$ in between the  energy equipartition exponent $1-r$ and the KZ exponent $2/3$, thereby altering the bounds for $ x_*$ compared to the 2D case.
The NLS system is the critical case $\alpha=2$, for which the Hopf value collides with the equipartition exponent $x_E=1-r$; see Fig.~\ref{fig:LimitCycle_final}. When approaching the NLS system from below, \emph{e.g.} $\alpha\to 2^-$, the isolated fixed point $P_+$  moves to $\infty$; see the green lines in Panels (a) of Fig.~\ref{fig:LimitCycle_final} and \ref{fig:LimitCycle_bridge}.
\end{enumerate}
\rroc

\corr
\paragraph{Basic classification of co-dimension 1 bifurcations.}
In both the generic and the non-generic cases previously described,
we expect the $\infty_*$-cycle to emerge out of a series of  co-dimension 1 bifurcations of limit cycles, which can be tracked down all the way either from infinity or from the Hopf point, provided the latter exists. Following the general classification of co-dimension 1 bifurcations \cite{kuznetsov2013elements}, those can either correspond to  \emph{flip},  \emph{fold} or  \emph{torus} bifurcations~\cite{kuznetsov2013elements}. As a very brief reminder, we here simply recall that a flip bifurcation of limit cycles  generically indicates a period doubling (PD), \emph{i.e.} a cycle emerges with twice the period of the original cycle, while  the original cycle  changes its stability. A fold bifurcation of limit cycles  generically corresponds to a turning point of a \emph{curve of limit cycles}, known as a limit point of cycle (LPC), and characterized by either the birth or the mutual annihilation of two cycles with different stability.  
The curve of limit cycles may be defined from any relevant feature of the cycles. Here, we  track either the minimal or the maximal value of the variable $f$ on the cycles with respect to $x$ \cite{kuznetsov2013elements}, as seen in Panels (a) of Fig. \ref{fig:LimitCycle_GW} \& \ref{fig:LimitCycle_NLS}.
Finally, the  torus (Neimark-Sacker) bifurcation of limit cycles  generically corresponds to a bifurcation of a cycle to an invariant torus, on which the flow contains periodic or quasi-periodic motions. 
\rroc


\corr
\rroc

\subsection{Generic and non-generic routes towards the $\infty_\star$-cycle.}
\corr
We now describe in more details the two bifurcation scenarios leading to the global bifurcation for the GW (generic) and the NLS (degenerate) cases. Unlike in 2D, where  there is one and only one limit cycle present for the values of $x$ in between of $x_H$ and $x_*$ (see Fig.~\ref{Fig:2d_conti} in 
 Appendix \ref{sec:GW_2}), a wider variety  of scenarios occurs in 4D, with  limiting cycles emerging or disappearing for values in between $x_H$ and $x_*$, or cycles  existing for for $x>x_*$. 

\begin{figure}
	\includegraphics[width=\columnwidth]{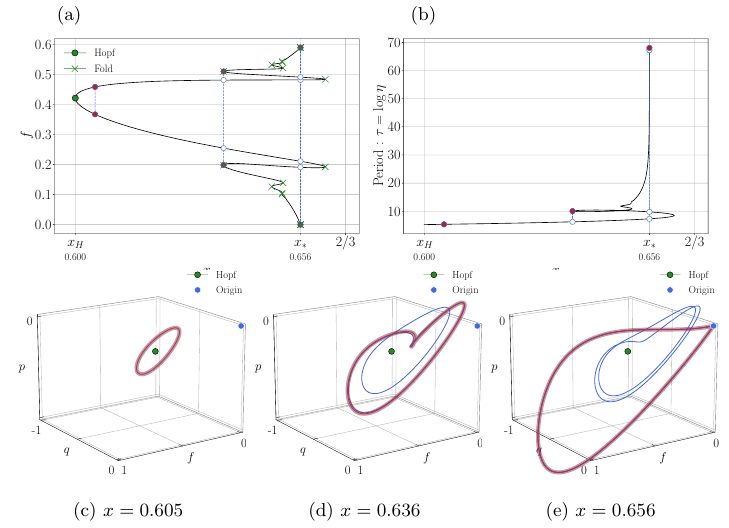}
	\caption{{\bf   Periodic solutions in the 3D GW case.} Panel (a) tracks the cycles by showing the minima and maxima of the  profiles $f(\tau)$. Panel (b)  shows the corresponding periods.  Panels (c,d,e) show 3D projections of the cycles found at different  $x$, indicated by the white and red dots in the previous panels (white dots--blue cycles, red dots--red cycles).  The $\infty_*$-cycle (in red) is found at  $x_*\simeq 0.656$.}
	\label{fig:LimitCycle_GW}
\end{figure}

\paragraph{Bifurcations  towards the GW $\infty_*$- cycle.}
  The curves presented in Fig.\ref{fig:LimitCycle_GW}   show the bifurcations of limit cycles leading to the $\infty_*$-cycle relevant in the GW case $\alpha=1$.  
  It is obtained by performing the numerical continuation initialized at $x = x_H \approx 0.600$,
  where a  (small) limit cycle emerges.
    Panel (a) tracks the  maximal and the minimal values of $f(\tau)$ encountered on the cycles when varying the parameter $x$. The corresponding periods are shown in Panel (b). 
     The leftmost (green) point in Panel (a) corresponds to the Hopf bifurcation, while  the pair of red dots closest at $x\approx 0.605$ features the birth of a small cycle; Its  3D projection onto the space $(f,p,q)$ is shown in Panel (c).    
  Increasing $x$, a fold bifurcation occurs at $x\approx 0.636$ giving birth to two more cycles; This event is signaled by the  second pair of  red dots on the vertical dashed line at $x\approx 0.636$ in Panel (a); As $x$ is increased, the two new cycles  separate from each other,  emerging out of the red cycle featured in Panel (d). 
  Note that the  cycle directly branching from the  Hopf bifurcation continues to exist: It is represented both by  the pair of white dots in Panel (a) at  $x\approx 0.636$, and   the blue curve in  Panel (d). Further increasing $x$,  four further fold bifurcations occur, leading to creation or annihilation of pairs of cycles. In particular, the cycles branching from the  Hopf point eventually vanishes by colliding with another remaining cycle at some $x\simeq 0.66 >x_*$, and  no cycles exist in the system for greater values of $x$. 
In this scenario, a unique  $\infty_*$- cycle appears at $x = x_* = 0.656$, and  is the red one in Panel (e). It is the outcome of a homoclinic bifurcation, with one of the cycle originating from the ultimate fold bifurcation colliding with the origin.
\rroc

\begin{figure}
	\includegraphics[width=\columnwidth]{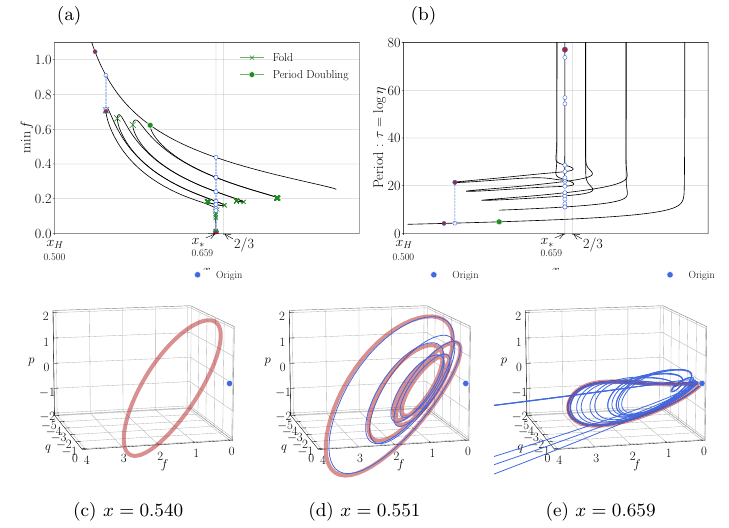}
	\caption{{\bf Periodic solutions in the 3D NLS case.} Panel (a) tracks the cycles by showing the minima of $f(\tau)$. Panel (b)  shows the corresponding periods.  Panels (c,d,e) show 3D projections of the cycles found at different values of $x$, indicated by the white and red dots in the previous panels.  The $\infty_*$-cycle shown in red in Panel (e) is found at  $x_*\simeq 0.659$.}
	\label{fig:LimitCycle_NLS}
\end{figure}

\corr
\paragraph{Bifurcations  towards the NLS $\infty_*$- cycle.}
\rroc
\corr
The 3D NLS system with $\alpha=2$ is a degenerate system for which the isolated fixed point $P_+$ has escaped to infinity.
The continuation algorithm then cannot be properly initiated from the Hopf point; To  construct the bifurcation  diagram at fixed $\alpha=2$, we rely on the Simpson strategy described in Appendix~\ref{sec:continuation}. In brief, the strategy consists in generating various co-dimension 2 continuation curves branching from  randomly selected points on the continuation curve $\alpha=1,x=x_1$ to $\alpha=2,x=x_2$. This  initializes  the continuation algorithms on various random branches on the plane $\alpha=2$, rather  than on the  degenerate Hopf point. 
The outcome of the process is represented in Fig.~\ref{fig:LimitCycle_NLS}.
\rroc
The bifurcation diagram is plotted in Panel~(a)  and the respective periods in Panel~(b) of Fig.\ref{fig:LimitCycle_NLS}. Panels~(c), (d) and (e) show the 3D projections of the cycles for the values of $x$ marked by the red and white dots in the previous panels. 
The Hopf cycle emanating from infinity corresponds to a periodic orbit of infinite amplitude in the  limit $x\to 0.5^+$.  From the Hopf point, the diagram features a sequence of  fold bifurcations, occurring at various values of  $x$. While more intricate, this scenario remains qualitatively similar to the generic example represented by the 3D GW case.
\corr
However, the NLS bifurcation scenario has also a series of distinctive features. 
First, as seen 
in Panel (d), the cycles emerging from the first three  fold bifurcations  have  rather intricate structure featuring up to seven loops; Panel (d) shows the already complicated pattern  at $x\simeq 0.551$ where three cycles exist together.
Second, in addition to the folds we  observe a flip bifurcation, represented by the green dots in Panels (a) and (b) at $x\approx 0.595$: The cycle branching  from the Hopf bifurcation then changes its  stability. Besides, Panel (b) shows clearly 
that the new cycle emerging  at $x\approx 0.595$ has  a period twice the original cycle marked by the green dot.
Third, in addition to $x_*$, there are four more values of $x$ at which the cycle periods turn very large (possibly infinite). 
These cycles are not $\infty_*$- cycles \corr because they  remain at finite distance from the origin \rroc, \emph{e.g} $f_{\min} $ remains finite, as explicit in Panel (a).  Still, such cycles are very large in the 4D phase space, with in particular $f_{\max} \to \infty$;
see also Panel (a) of Fig.~\ref{fig:LimitCycle_bridge}.

A closer inspection (not shown here) reveals that pieces of \corr those spurious  giant cycles follow the line $(f, -f/2, -f/3, -f)$; On this line, the 4D system satisfies 
$\dot f=\dot p=\dot q = 0$ and $\dot g=-2/(3f) \to 0$ as $f \to \infty$,
meaning that the giant cycles indicate  degenerate homoclinic bifurcations, featuring the collision of a cycle with a fixed point $P_+$ at infinity.
\corr
Fourth, unlike the GW case, there exists a great, possibly infinite, number of folds occurring in the small vicinity of the 
value $x =x_* \approx 0.659$ for which
the $\infty_*$- cycle emerges; This is signaled in Panel (e) by the numerous cycles in blues, apparently arbitrarily  close to the $\infty_*$-cycle (red).
\corr
As an aside, we note that for the cases $\alpha>2$ or equivalently $r<1/2$, the point  $P_+$ does not exist, but the bifurcation diagram is qualitatively similar to the one of the singular 3D NLS case with $\alpha=2$. In particular, the giant cycles containing  straight line segments continue to exist, as well as the very large number of folds arbitrarily close to $x_*$.

\rroc


\subsection{The $\infty_*$- cycle in the general case: observations and conjectures.}
\begin{figure}
	\includegraphics[width=\columnwidth]{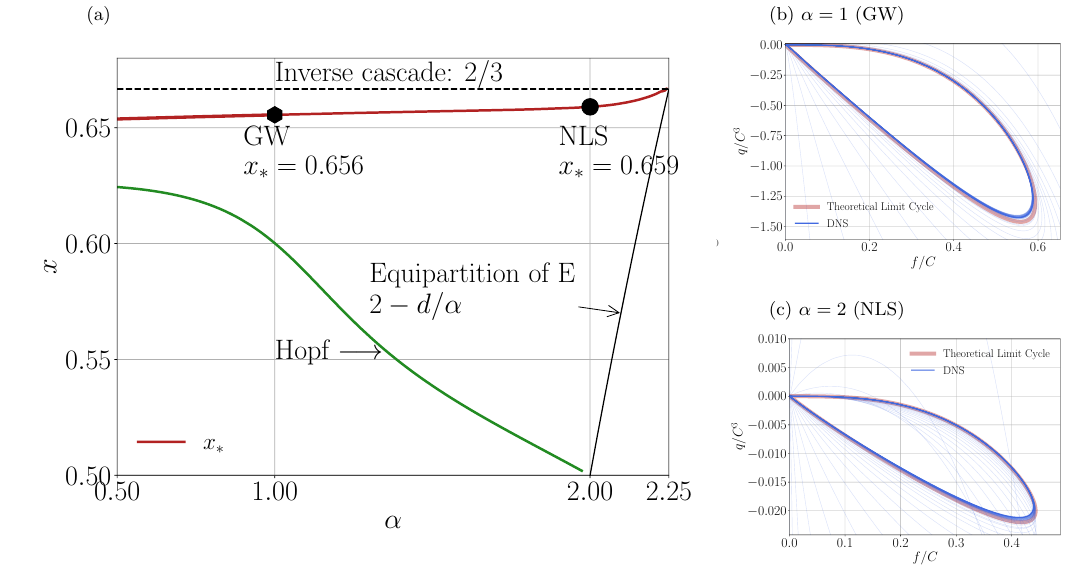}
	\caption{{\bf Full phase portrait}. Panel (a) shows the various scaling exponents as a function of $\alpha$ and fixed $d=3$.
	The  line of anomalous exponent $x_*$ is the practical outcome of our analysis. Panels (b,c) compare the $\infty_*$-cycles with the DNS profiles reported in Fig.\ref{fig:DNS_compensated}. The fine blue lines are transients before the simulation adjusts to self-similar profile of the second-kind. \corr See also videos online \rroc.} 
		\label{fig:LimitCycle_final}
\end{figure}

\corr
To reveal general properties of the self-similar solutions of the second kind for both generic and non-generic fourth-order systems,  we now compute the continuation curve  varying parameter $\alpha$ from $0.50$ to $2.25$ and $x$ from $0.5$ to $2.25$, hereby directly tracking  the $\infty_*$- cycles. In practice, this co-dimension 2 continuation is initialized from the $\infty_*$- cycles at $\alpha=1$.
Our findings are summarised in Fig.~\ref{fig:LimitCycle_final}. Panel (a) shows the behavior of the exponents $x_H$ and  $x_*$ as a function of $\alpha$, comparing it to the energy equipartition exponent $x_E(\alpha)=2-d/\alpha$, as well as the KZ of exponent $x_Q=2/3$ for the stationary inverse cascade KZ spectrum. Note that the the Hopf point and the energy equipartition coincide at $\alpha=2$, \emph{e.g} $x_H = x_E$; At $\alpha=2.25$, three exponents coincide, namely $x_E = x_Q=x_*$.
Panels (b) and (c) show 2D projections on which the  $\infty_*$- cycles  are overlaid with the orbits obtained by  numerical simulations of the respective fourth-order  differential equation showing the evolving spectra arising  for finite-support initial data. Agreement between the profiles is excellent, and this confirms that the value of $x_*$ determined by numerical continuation is indeed the one relevant for the second-kind self-similarity.
\rroc


Based on the information presented 
in~Fig.\ref{fig:LimitCycle_final}, we  propose
the following conjectures:
\begin{itemize}
	\item[(i)]  $x_*$ exists and is unique for all $r>1/3$;
	\item[(ii)] $x_*$ is bounded: $\min(x_H,x_E=1-r)<x_*
	<x_Q= 2/3 $;
	\item[(iii)] $x_* \to x_Q = 2/3$ as $r\to 1/3$. In this limit, the energy equipartition exponent $x_E$
	coincides with the one of  inverse cascade KZ, $x_Q$. This should correspond to the finite-time blow-up becoming infinite, {\em i.e.} $t_* \to \infty$;
	\item[(iv)] The $\infty_*$-cycles represent stable self-similar solutions of the second kind which are attractors for the spectra evolving out of finite-support initial data of arbitrary shape.
\end{itemize}

\section{Concluding remarks}
\label{sec:conclusion}
	\corr Using  numerical simulations  and a self-similarity Ansatz \rroc, we  have shown the existence of anomalous transients in the inverse cascade of fourth-order DAM~(\ref{eq:LeithFour}) relevant for a  class of wave-turbulence systems.
	We focused on a continuous one-parametric class of systems which includes   the 3D GW and  the 3D NLS as special cases. Such anomalous transients are mis-characterized in second-order DAM, which fail to distinguish between  the 3D GW and the 3D NLS cases. By taking into account the correct interplay between KL and KZ solutions, the fourth-order DAM are found to feature weak but non-trivial systematic deviations between the anomalous transients and the KZ solutions. 	\corr From  careful  numerical simulations, we have identified the presence of  anomalous scaling, and observed that the  deviations to KZ scaling  are very small, less than $1\%$ for both the 3D GW and the 3D NLS cases.\rroc
	Those transients can be precisely related to convergence of the solutions to  self-similar solution of the second-kind featuring finite-time blow-up of the propagating front, and this is our main result. This characterization allows for a very precise determination of the exponent and of the self-similar profile using the theory of dynamical systems.
		We provided a systematic way of determining the anomalous transients. To that end, we extended and reformulated a previous analysis of self-similarity of the second kind in second-order DAM, to relate $x_*$ to the existence of an $\infty_*$-cycle. Rather than direct computations of trajectories of the associated dynamical system,  \emph{e.g.} using shooting methods which in the present case prove highly inefficient, we relied on numerical continuation software, to chase and identify the cycle in the 4D phase space.  This  is, to our knowledge, the only  efficient way of determining  the exponent $ x_*$.	Agreement with the  numerical simulations  is found to be  excellent, proving that the  found self-similar solutions are stable: They represent the large-time asymptotics for spectra evolving out of  finite-support initial data of arbitrary shape.
		Our findings about the self-similar transients are summarised as a set of four conjectures in the end of the previous section.
		
		Among natural perspectives for this work is the full classification of  the self-similar transients in a wider class of fourth-order DAM given by Eq.~(\ref{eq:LeithFour_gen}),  as well as extending the description by including the finite-capacity  direct cascades~\cite{connaughton2003non}.
		In particular, it would be useful to include the important example of the gravity wave turbulence on the deep water surface.
		Even more challenging but important task would be to extend our approach to the integro-differential wave-kinetic equations which are more realistic in describing wave turbulence than the differential models considered in the present article. 
		\corr In the same spirit, it would also be interesting to study other integro-differential closures, possibly including turbulent systems with spectra varying in both the wave number and the physical space, \emph{e.g.},   non-Markovian Liouville equation suggested in Refs.\cite{Balescu1,Balescu2} for the plasma drift waves.\rroc

\section*{acknowledgements}	
The authors thank A. Mailybaev for useful discussions.
ST acknowledges support from the  Programa de Capacitação Institucional   of CNPq and the French-Brazilian network in mathematics. The work of VG was partially supported by the ``chercheurs invit\'es" awards of
the F\'ed\'eration Doeblin FR 2800, Universit\'e de la C\^ote d'Azur, France.  The work of SN was supported by the Chaire  D'Excellence IDEX (Initiative of Excellence) awarded by
Universit\'e de la C\^ote d'Azur, France,   Simons  Foundation Collaboration grant Wave Turbulence (Award ID 651471), the  European  Unions  Horizon  2020 research and innovation programme  in the framework of Marie Skodowska-Curie HALT project (grant agreement No 823937) and  the FET Flagships PhoQuS project
(grant agreement No 820392). The work of SM was supported by state funding program FSUS-2020-0034.

\bibliography{biblio}

\begin{thebibliography}{41}%
\makeatletter
\providecommand \@ifxundefined [1]{%
 \@ifx{#1\undefined}
}%
\providecommand \@ifnum [1]{%
 \ifnum #1\expandafter \@firstoftwo
 \else \expandafter \@secondoftwo
 \fi
}%
\providecommand \@ifx [1]{%
 \ifx #1\expandafter \@firstoftwo
 \else \expandafter \@secondoftwo
 \fi
}%
\providecommand \natexlab [1]{#1}%
\providecommand \enquote  [1]{``#1''}%
\providecommand \bibnamefont  [1]{#1}%
\providecommand \bibfnamefont [1]{#1}%
\providecommand \citenamefont [1]{#1}%
\providecommand \href@noop [0]{\@secondoftwo}%
\providecommand \href [0]{\begingroup \@sanitize@url \@href}%
\providecommand \@href[1]{\@@startlink{#1}\@@href}%
\providecommand \@@href[1]{\endgroup#1\@@endlink}%
\providecommand \@sanitize@url [0]{\catcode `\\12\catcode `\$12\catcode
  `\&12\catcode `\#12\catcode `\^12\catcode `\_12\catcode `\%12\relax}%
\providecommand \@@startlink[1]{}%
\providecommand \@@endlink[0]{}%
\providecommand \url  [0]{\begingroup\@sanitize@url \@url }%
\providecommand \@url [1]{\endgroup\@href {#1}{\urlprefix }}%
\providecommand \urlprefix  [0]{URL }%
\providecommand \Eprint [0]{\href }%
\providecommand \doibase [0]{http://dx.doi.org/}%
\providecommand \selectlanguage [0]{\@gobble}%
\providecommand \bibinfo  [0]{\@secondoftwo}%
\providecommand \bibfield  [0]{\@secondoftwo}%
\providecommand \translation [1]{[#1]}%
\providecommand \BibitemOpen [0]{}%
\providecommand \bibitemStop [0]{}%
\providecommand \bibitemNoStop [0]{.\EOS\space}%
\providecommand \EOS [0]{\spacefactor3000\relax}%
\providecommand \BibitemShut  [1]{\csname bibitem#1\endcsname}%
\let\auto@bib@innerbib\@empty
\bibitem [{\citenamefont {Leith}(1967)}]{leith1967diffusion}%
  \BibitemOpen
  \bibfield  {author} {\bibinfo {author} {\bibfnamefont {C.}~\bibnamefont
  {Leith}},\ }\bibfield  {title} {\enquote {\bibinfo {title} {Diffusion
  approximation to inertial energy transfer in isotropic turbulence},}\
  }\href@noop {} {\bibfield  {journal} {\bibinfo  {journal} {The Physics of
  Fluids}\ }\textbf {\bibinfo {volume} {10}},\ \bibinfo {pages} {1409--1416}
  (\bibinfo {year} {1967})}\BibitemShut {NoStop}%
\bibitem [{\citenamefont {Lee}(1952)}]{lee1952some}%
  \BibitemOpen
  \bibfield  {author} {\bibinfo {author} {\bibfnamefont {T.}~\bibnamefont
  {Lee}},\ }\bibfield  {title} {\enquote {\bibinfo {title} {On some statistical
  properties of hydrodynamical and magneto-hydrodynamical fields},}\
  }\href@noop {} {\bibfield  {journal} {\bibinfo  {journal} {Quarterly of
  Applied Mathematics}\ }\textbf {\bibinfo {volume} {10}},\ \bibinfo {pages}
  {69--74} (\bibinfo {year} {1952})}\BibitemShut {NoStop}%
\bibitem [{\citenamefont {Rubinstein}, \citenamefont {Clark},\ and\
  \citenamefont {Kurien}(2017)}]{rubinstein2017leith}%
  \BibitemOpen
  \bibfield  {author} {\bibinfo {author} {\bibfnamefont {R.}~\bibnamefont
  {Rubinstein}}, \bibinfo {author} {\bibfnamefont {T.}~\bibnamefont {Clark}}, \
  and\ \bibinfo {author} {\bibfnamefont {S.}~\bibnamefont {Kurien}},\
  }\bibfield  {title} {\enquote {\bibinfo {title} {Leith diffusion model for
  homogeneous anisotropic turbulence},}\ }\href@noop {} {\bibfield  {journal}
  {\bibinfo  {journal} {Computers \& Fluids}\ }\textbf {\bibinfo {volume}
  {151}},\ \bibinfo {pages} {108--114} (\bibinfo {year} {2017})}\BibitemShut
  {NoStop}%
\bibitem [{\citenamefont {Clark}, \citenamefont {Rubinstein},\ and\
  \citenamefont {Weinstock}(2009)}]{clark2009reassessment}%
  \BibitemOpen
  \bibfield  {author} {\bibinfo {author} {\bibfnamefont {T.}~\bibnamefont
  {Clark}}, \bibinfo {author} {\bibfnamefont {R.}~\bibnamefont {Rubinstein}}, \
  and\ \bibinfo {author} {\bibfnamefont {J.}~\bibnamefont {Weinstock}},\
  }\bibfield  {title} {\enquote {\bibinfo {title} {Reassessment of the
  classical turbulence closures: the {L}eith diffusion model},}\ }\href@noop {}
  {\bibfield  {journal} {\bibinfo  {journal} {Journal of Turbulence}\ ,\
  \bibinfo {pages} {N35}} (\bibinfo {year} {2009})}\BibitemShut {NoStop}%
\bibitem [{\citenamefont {Orszag}\ and\ \citenamefont
  {Raila}(1973)}]{orszag1973test}%
  \BibitemOpen
  \bibfield  {author} {\bibinfo {author} {\bibfnamefont {S.}~\bibnamefont
  {Orszag}}\ and\ \bibinfo {author} {\bibfnamefont {D.}~\bibnamefont {Raila}},\
  }\bibfield  {title} {\enquote {\bibinfo {title} {Test of spectral energy
  transfer models of turbulence decay},}\ }\href@noop {} {\bibfield  {journal}
  {\bibinfo  {journal} {The Physics of Fluids}\ }\textbf {\bibinfo {volume}
  {16}},\ \bibinfo {pages} {172--173} (\bibinfo {year} {1973})}\BibitemShut
  {NoStop}%
\bibitem [{\citenamefont {Connaughton}\ and\ \citenamefont
  {Nazarenko}(2004)}]{connaughton2004warm}%
  \BibitemOpen
  \bibfield  {author} {\bibinfo {author} {\bibfnamefont {C.}~\bibnamefont
  {Connaughton}}\ and\ \bibinfo {author} {\bibfnamefont {S.}~\bibnamefont
  {Nazarenko}},\ }\bibfield  {title} {\enquote {\bibinfo {title} {Warm cascades
  and anomalous scaling in a diffusion model of turbulence},}\ }\href@noop {}
  {\bibfield  {journal} {\bibinfo  {journal} {Physical review letters}\
  }\textbf {\bibinfo {volume} {92}},\ \bibinfo {pages} {044501} (\bibinfo
  {year} {2004})}\BibitemShut {NoStop}%
\bibitem [{\citenamefont {Lilly}(1989)}]{Lilly}%
  \BibitemOpen
  \bibfield  {author} {\bibinfo {author} {\bibfnamefont {D.~K.}\ \bibnamefont
  {Lilly}},\ }\bibfield  {title} {\enquote {\bibinfo {title} {Two-dimensional
  turbulence generated by energy sources at two scales},}\ }\href {\doibase
  10.1175/1520-0469(1989)046<2026:TDTGBE>2.0.CO;2} {\bibfield  {journal}
  {\bibinfo  {journal} {Journal of Atmospheric Sciences}\ }\textbf {\bibinfo
  {volume} {46}},\ \bibinfo {pages} {2026 -- 2030} (\bibinfo {year}
  {1989})}\BibitemShut {NoStop}%
\bibitem [{\citenamefont {Morel}, \citenamefont {Xu},\ and\ \citenamefont
  {Gurcan}(2021)}]{Morel}%
  \BibitemOpen
  \bibfield  {author} {\bibinfo {author} {\bibfnamefont {P.}~\bibnamefont
  {Morel}}, \bibinfo {author} {\bibfnamefont {S.}~\bibnamefont {Xu}}, \ and\
  \bibinfo {author} {\bibfnamefont {O.~D.}\ \bibnamefont {Gurcan}},\ }\bibfield
   {title} {\enquote {\bibinfo {title} {A differential approximation model for
  passive scalar turbulence},}\ }\href
  {https://doi.org/10.1088/1751-8121/ac1484} {\bibfield  {journal} {\bibinfo
  {journal} {Journal of Physics A: Mathematical and Theoretical}\ }\textbf
  {\bibinfo {volume} {54}},\ \bibinfo {pages} {335701} (\bibinfo {year}
  {2021})}\BibitemShut {NoStop}%
\bibitem [{\citenamefont {L’vov}, \citenamefont {Nazarenko},\ and\
  \citenamefont {Volovik}(2004)}]{LvovVolovik}%
  \BibitemOpen
  \bibfield  {author} {\bibinfo {author} {\bibfnamefont {V.}~\bibnamefont
  {L’vov}}, \bibinfo {author} {\bibfnamefont {S.~V.}\ \bibnamefont
  {Nazarenko}}, \ and\ \bibinfo {author} {\bibfnamefont {G.}~\bibnamefont
  {Volovik}},\ }\bibfield  {title} {\enquote {\bibinfo {title} {Energy spectra
  of developed superfluid turbulence},}\ }\href
  {https://doi.org/10.1134/1.1839294} {\bibfield  {journal} {\bibinfo
  {journal} {Jetp Lett.}\ }\textbf {\bibinfo {volume} {80}},\ \bibinfo {pages}
  {479–483} (\bibinfo {year} {2004})}\BibitemShut {NoStop}%
\bibitem [{\citenamefont {L’vov}, \citenamefont {Nazarenko},\ and\
  \citenamefont {Skrbek}(2006)}]{LvovSkrbek}%
  \BibitemOpen
  \bibfield  {author} {\bibinfo {author} {\bibfnamefont {V.}~\bibnamefont
  {L’vov}}, \bibinfo {author} {\bibfnamefont {S.~V.}\ \bibnamefont
  {Nazarenko}}, \ and\ \bibinfo {author} {\bibfnamefont {L.}~\bibnamefont
  {Skrbek}},\ }\bibfield  {title} {\enquote {\bibinfo {title} {Energy spectra
  of developed turbulence in helium superfluids},}\ }\href
  {https://doi.org/10.1007/s10909-006-9230-8} {\bibfield  {journal} {\bibinfo
  {journal} {J Low Temp Phys}\ }\textbf {\bibinfo {volume} {145}},\ \bibinfo
  {pages} {125–142} (\bibinfo {year} {2006})}\BibitemShut {NoStop}%
\bibitem [{\citenamefont {Thalabard}\ \emph {et~al.}(2015)\citenamefont
  {Thalabard}, \citenamefont {Nazarenko}, \citenamefont {Galtier},\ and\
  \citenamefont {Medvedev}}]{thalabard2015anomalous}%
  \BibitemOpen
  \bibfield  {author} {\bibinfo {author} {\bibfnamefont {S.}~\bibnamefont
  {Thalabard}}, \bibinfo {author} {\bibfnamefont {S.}~\bibnamefont
  {Nazarenko}}, \bibinfo {author} {\bibfnamefont {S.}~\bibnamefont {Galtier}},
  \ and\ \bibinfo {author} {\bibfnamefont {S.}~\bibnamefont {Medvedev}},\
  }\bibfield  {title} {\enquote {\bibinfo {title} {Anomalous spectral laws in
  differential models of turbulence},}\ }\href@noop {} {\bibfield  {journal}
  {\bibinfo  {journal} {Journal of Physics A: Mathematical and Theoretical}\
  }\textbf {\bibinfo {volume} {48}},\ \bibinfo {pages} {285501} (\bibinfo
  {year} {2015})}\BibitemShut {NoStop}%
\bibitem [{\citenamefont {Galtier}, \citenamefont {Nazarenko},\ and\
  \citenamefont {Thalabard}(2019)}]{galtier2019nonlinear}%
  \BibitemOpen
  \bibfield  {author} {\bibinfo {author} {\bibfnamefont {S.}~\bibnamefont
  {Galtier}}, \bibinfo {author} {\bibfnamefont {E.}~\bibnamefont {Nazarenko},
  \bibfnamefont {S.and~Buchlin}}, \ and\ \bibinfo {author} {\bibfnamefont
  {S.}~\bibnamefont {Thalabard}},\ }\bibfield  {title} {\enquote {\bibinfo
  {title} {Nonlinear diffusion models for gravitational wave turbulence},}\
  }\href@noop {} {\bibfield  {journal} {\bibinfo  {journal} {Physica D:
  Nonlinear Phenomena}\ }\textbf {\bibinfo {volume} {390}},\ \bibinfo {pages}
  {84--88} (\bibinfo {year} {2019})}\BibitemShut {NoStop}%
\bibitem [{\citenamefont {L’vov}\ \emph {et~al.}(1998)\citenamefont
  {L’vov}, \citenamefont {Podivilov}, \citenamefont {Pomyalov}, \citenamefont
  {Procaccia},\ and\ \citenamefont {Vandembroucq}}]{l1998improved}%
  \BibitemOpen
  \bibfield  {author} {\bibinfo {author} {\bibfnamefont {V.}~\bibnamefont
  {L’vov}}, \bibinfo {author} {\bibfnamefont {E.}~\bibnamefont {Podivilov}},
  \bibinfo {author} {\bibfnamefont {A.}~\bibnamefont {Pomyalov}}, \bibinfo
  {author} {\bibfnamefont {I.}~\bibnamefont {Procaccia}}, \ and\ \bibinfo
  {author} {\bibfnamefont {D.}~\bibnamefont {Vandembroucq}},\ }\bibfield
  {title} {\enquote {\bibinfo {title} {Improved shell model of turbulence},}\
  }\href@noop {} {\bibfield  {journal} {\bibinfo  {journal} {Physical Review
  E}\ }\textbf {\bibinfo {volume} {58}},\ \bibinfo {pages} {1811} (\bibinfo
  {year} {1998})}\BibitemShut {NoStop}%
\bibitem [{\citenamefont {Campolina}\ and\ \citenamefont
  {Mailybaev}(2018)}]{campolina2018chaotic}%
  \BibitemOpen
  \bibfield  {author} {\bibinfo {author} {\bibfnamefont {C.}~\bibnamefont
  {Campolina}}\ and\ \bibinfo {author} {\bibfnamefont {A.}~\bibnamefont
  {Mailybaev}},\ }\bibfield  {title} {\enquote {\bibinfo {title} {Chaotic
  blowup in the {3D} incompressible {E}uler equations on a logarithmic
  lattice},}\ }\href@noop {} {\bibfield  {journal} {\bibinfo  {journal}
  {Physical review letters}\ }\textbf {\bibinfo {volume} {121}},\ \bibinfo
  {pages} {064501} (\bibinfo {year} {2018})}\BibitemShut {NoStop}%
\bibitem [{\citenamefont {Grebenev}\ \emph {et~al.}(2014)\citenamefont
  {Grebenev}, \citenamefont {Nazarenko}, \citenamefont {Medvedev},
  \citenamefont {Chirkunov},\ and\ \citenamefont {Schwab}}]{GNMCh2014}%
  \BibitemOpen
  \bibfield  {author} {\bibinfo {author} {\bibfnamefont {V.}~\bibnamefont
  {Grebenev}}, \bibinfo {author} {\bibfnamefont {S.}~\bibnamefont {Nazarenko}},
  \bibinfo {author} {\bibfnamefont {S.}~\bibnamefont {Medvedev}}, \bibinfo
  {author} {\bibfnamefont {Y.}~\bibnamefont {Chirkunov}}, \ and\ \bibinfo
  {author} {\bibfnamefont {I.}~\bibnamefont {Schwab}},\ }\bibfield  {title}
  {\enquote {\bibinfo {title} {Self-similar solution in {Leith} model of
  turbulence: anomalous power law and asymptotic analysis},}\ }\href@noop {}
  {\bibfield  {journal} {\bibinfo  {journal} {Journal of Physics A:
  Mathematical and Theoretical}\ }\textbf {\bibinfo {volume} {47}},\ \bibinfo
  {pages} {025401} (\bibinfo {year} {2014})}\BibitemShut {NoStop}%
\bibitem [{\citenamefont {Nazarenko}(2011)}]{nazarenko2011wave}%
  \BibitemOpen
  \bibfield  {author} {\bibinfo {author} {\bibfnamefont {S.}~\bibnamefont
  {Nazarenko}},\ }\href@noop {} {\emph {\bibinfo {title} {Wave turbulence}}},\
  Vol.\ \bibinfo {volume} {825}\ (\bibinfo  {publisher} {Springer Science \&
  Business Media},\ \bibinfo {year} {2011})\BibitemShut {NoStop}%
\bibitem [{\citenamefont {L’vov}\ and\ \citenamefont
  {Nazarenko}(2006)}]{l2006differential}%
  \BibitemOpen
  \bibfield  {author} {\bibinfo {author} {\bibfnamefont {V.}~\bibnamefont
  {L’vov}}\ and\ \bibinfo {author} {\bibfnamefont {S.}~\bibnamefont
  {Nazarenko}},\ }\bibfield  {title} {\enquote {\bibinfo {title} {Differential
  model for {2D} turbulence},}\ }\href@noop {} {\bibfield  {journal} {\bibinfo
  {journal} {JETP letters}\ }\textbf {\bibinfo {volume} {83}},\ \bibinfo
  {pages} {541--545} (\bibinfo {year} {2006})}\BibitemShut {NoStop}%
\bibitem [{\citenamefont {Hasselmann}\ \emph {et~al.}(1985)\citenamefont
  {Hasselmann}, \citenamefont {Hasselmann}, \citenamefont {Allender},\ and\
  \citenamefont {Barnett}}]{Hasselmann}%
  \BibitemOpen
  \bibfield  {author} {\bibinfo {author} {\bibfnamefont {S.}~\bibnamefont
  {Hasselmann}}, \bibinfo {author} {\bibfnamefont {K.}~\bibnamefont
  {Hasselmann}}, \bibinfo {author} {\bibfnamefont {J.~H.}\ \bibnamefont
  {Allender}}, \ and\ \bibinfo {author} {\bibfnamefont {T.~P.}\ \bibnamefont
  {Barnett}},\ }\bibfield  {title} {\enquote {\bibinfo {title} {Computations
  and parameterizations of the nonlinear energy transfer in a gravity-wave
  specturm. part ii: Parameterizations of the nonlinear energy transfer for
  application in wave models},}\ }\href {\doibase
  10.1175/1520-0485(1985)015<1378:CAPOTN>2.0.CO;2} {\bibfield  {journal}
  {\bibinfo  {journal} {Journal of Physical Oceanography}\ }\textbf {\bibinfo
  {volume} {15}},\ \bibinfo {pages} {1378 -- 1391} (\bibinfo {year}
  {1985})}\BibitemShut {NoStop}%
\bibitem [{\citenamefont {Nazarenko}(2006)}]{DAMkelvin}%
  \BibitemOpen
  \bibfield  {author} {\bibinfo {author} {\bibfnamefont {S.}~\bibnamefont
  {Nazarenko}},\ }\bibfield  {title} {\enquote {\bibinfo {title} {Differential
  approximation for kelvin wave turbulence},}\ }\href
  {https://doi.org/10.1134/S0021364006050031} {\bibfield  {journal} {\bibinfo
  {journal} {Jetp Lett.}\ }\textbf {\bibinfo {volume} {83}},\ \bibinfo {pages}
  {198–200} (\bibinfo {year} {2006})}\BibitemShut {NoStop}%
\bibitem [{\citenamefont {Galtier}\ and\ \citenamefont
  {Nazarenko}(2017)}]{galtier2017turbulence}%
  \BibitemOpen
  \bibfield  {author} {\bibinfo {author} {\bibfnamefont {S.}~\bibnamefont
  {Galtier}}\ and\ \bibinfo {author} {\bibfnamefont {S.}~\bibnamefont
  {Nazarenko}},\ }\bibfield  {title} {\enquote {\bibinfo {title} {Turbulence of
  weak gravitational waves in the early universe},}\ }\href@noop {} {\bibfield
  {journal} {\bibinfo  {journal} {Physical Review Letters}\ }\textbf {\bibinfo
  {volume} {119}},\ \bibinfo {pages} {221101} (\bibinfo {year}
  {2017})}\BibitemShut {NoStop}%
\bibitem [{\citenamefont {Skipp}\ and\ \citenamefont
  {Nazarenko}(2020)}]{skipp2020wave}%
  \BibitemOpen
  \bibfield  {author} {\bibinfo {author} {\bibfnamefont {V.}~\bibnamefont
  {Skipp}, \bibfnamefont {J.and~L'vov}}\ and\ \bibinfo {author} {\bibfnamefont
  {S.}~\bibnamefont {Nazarenko}},\ }\bibfield  {title} {\enquote {\bibinfo
  {title} {Wave turbulence in self-gravitating {B}ose gases and nonlocal
  nonlinear optics},}\ }\href@noop {} {\bibfield  {journal} {\bibinfo
  {journal} {Physical Review A}\ }\textbf {\bibinfo {volume} {102}},\ \bibinfo
  {pages} {043318} (\bibinfo {year} {2020})}\BibitemShut {NoStop}%
\bibitem [{\citenamefont {Connaughton}\ and\ \citenamefont
  {Pomeau}(2004)}]{connaughton2004kinetic}%
  \BibitemOpen
  \bibfield  {author} {\bibinfo {author} {\bibfnamefont {C.}~\bibnamefont
  {Connaughton}}\ and\ \bibinfo {author} {\bibfnamefont {Y.}~\bibnamefont
  {Pomeau}},\ }\bibfield  {title} {\enquote {\bibinfo {title} {Kinetic theory
  and {B}ose--{E}instein condensation},}\ }\href@noop {} {\bibfield  {journal}
  {\bibinfo  {journal} {Comptes Rendus Physique}\ }\textbf {\bibinfo {volume}
  {5}},\ \bibinfo {pages} {91--106} (\bibinfo {year} {2004})}\BibitemShut
  {NoStop}%
\bibitem [{\citenamefont {Nazarenko}\ and\ \citenamefont
  {Grebenev}(2016)}]{nazarenko2016self}%
  \BibitemOpen
  \bibfield  {author} {\bibinfo {author} {\bibfnamefont {S.}~\bibnamefont
  {Nazarenko}}\ and\ \bibinfo {author} {\bibfnamefont {V.}~\bibnamefont
  {Grebenev}},\ }\bibfield  {title} {\enquote {\bibinfo {title} {Self-similar
  formation of the {Kolmogorov} spectrum in the {Leith} model of turbulence},}\
  }\href@noop {} {\bibfield  {journal} {\bibinfo  {journal} {Journal of Physics
  A: Mathematical and Theoretical}\ }\textbf {\bibinfo {volume} {50}},\
  \bibinfo {pages} {035501} (\bibinfo {year} {2016})}\BibitemShut {NoStop}%
\bibitem [{\citenamefont {Leith}(1968)}]{leith1968diffusion}%
  \BibitemOpen
  \bibfield  {author} {\bibinfo {author} {\bibfnamefont {C.}~\bibnamefont
  {Leith}},\ }\bibfield  {title} {\enquote {\bibinfo {title} {Diffusion
  approximation for two-dimensional turbulence},}\ }\href@noop {} {\bibfield
  {journal} {\bibinfo  {journal} {The Physics of Fluids}\ }\textbf {\bibinfo
  {volume} {11}},\ \bibinfo {pages} {671--672} (\bibinfo {year}
  {1968})}\BibitemShut {NoStop}%
\bibitem [{\citenamefont {Kraichnan}(1967)}]{kraichnan1967inertial}%
  \BibitemOpen
  \bibfield  {author} {\bibinfo {author} {\bibfnamefont {R.}~\bibnamefont
  {Kraichnan}},\ }\bibfield  {title} {\enquote {\bibinfo {title} {Inertial
  ranges in two-dimensional turbulence},}\ }\href@noop {} {\bibfield  {journal}
  {\bibinfo  {journal} {The Physics of Fluids}\ }\textbf {\bibinfo {volume}
  {10}},\ \bibinfo {pages} {1417--1423} (\bibinfo {year} {1967})}\BibitemShut
  {NoStop}%
\bibitem [{\citenamefont {Nazarenko}\ \emph {et~al.}(2019)\citenamefont
  {Nazarenko}, \citenamefont {Grebenev}, \citenamefont {Medvedev},\ and\
  \citenamefont {Galtier}}]{nazarenko2019focusing}%
  \BibitemOpen
  \bibfield  {author} {\bibinfo {author} {\bibfnamefont {S.}~\bibnamefont
  {Nazarenko}}, \bibinfo {author} {\bibfnamefont {V.}~\bibnamefont {Grebenev}},
  \bibinfo {author} {\bibfnamefont {S.}~\bibnamefont {Medvedev}}, \ and\
  \bibinfo {author} {\bibfnamefont {S.}~\bibnamefont {Galtier}},\ }\bibfield
  {title} {\enquote {\bibinfo {title} {The focusing problem for the {L}eith
  model of turbulence: a self-similar solution of the third kind},}\
  }\href@noop {} {\bibfield  {journal} {\bibinfo  {journal} {Journal of Physics
  A: Mathematical and Theoretical}\ }\textbf {\bibinfo {volume} {52}},\
  \bibinfo {pages} {155501} (\bibinfo {year} {2019})}\BibitemShut {NoStop}%
\bibitem [{\citenamefont {Bell}\ \emph {et~al.}(2017)\citenamefont {Bell},
  \citenamefont {Grebenev}, \citenamefont {Medvedev},\ and\ \citenamefont
  {Nazarenko}}]{bell2017self}%
  \BibitemOpen
  \bibfield  {author} {\bibinfo {author} {\bibfnamefont {N.}~\bibnamefont
  {Bell}}, \bibinfo {author} {\bibfnamefont {V.}~\bibnamefont {Grebenev}},
  \bibinfo {author} {\bibfnamefont {S.}~\bibnamefont {Medvedev}}, \ and\
  \bibinfo {author} {\bibfnamefont {S.}~\bibnamefont {Nazarenko}},\ }\bibfield
  {title} {\enquote {\bibinfo {title} {Self-similar evolution of alfven wave
  turbulence},}\ }\href@noop {} {\bibfield  {journal} {\bibinfo  {journal}
  {Journal of Physics A: Mathematical and Theoretical}\ }\textbf {\bibinfo
  {volume} {50}},\ \bibinfo {pages} {435501} (\bibinfo {year}
  {2017})}\BibitemShut {NoStop}%
\bibitem [{\citenamefont {Semisalov}\ \emph {et~al.}(2021)\citenamefont
  {Semisalov}, \citenamefont {Grebenev}, \citenamefont {Medvedev},\ and\
  \citenamefont {Nazarenko}}]{SEMISALOV2021105903}%
  \BibitemOpen
  \bibfield  {author} {\bibinfo {author} {\bibfnamefont {B.}~\bibnamefont
  {Semisalov}}, \bibinfo {author} {\bibfnamefont {V.}~\bibnamefont {Grebenev}},
  \bibinfo {author} {\bibfnamefont {S.}~\bibnamefont {Medvedev}}, \ and\
  \bibinfo {author} {\bibfnamefont {S.}~\bibnamefont {Nazarenko}},\ }\bibfield
  {title} {\enquote {\bibinfo {title} {Numerical analysis of a self-similar
  turbulent flow in bose–einstein condensates},}\ }\href {\doibase
  https://doi.org/10.1016/j.cnsns.2021.105903} {\bibfield  {journal} {\bibinfo
  {journal} {Communications in Nonlinear Science and Numerical Simulation}\
  }\textbf {\bibinfo {volume} {102}},\ \bibinfo {pages} {105903} (\bibinfo
  {year} {2021})}\BibitemShut {NoStop}%
\bibitem [{Note1()}]{Note1}%
  \BibitemOpen
  \bibinfo {note} {In Appendix~\ref {sec:GW_2}, we show that upon using
  variables akin to $f,g,p,q$ variables which we introduced to address the
  fourth-order DAM, the global bifurcation which determines $ x_*$ for the
  second-order DAM describes the appearance of a homoclinic cycle.}\BibitemShut
  {Stop}%
\bibitem [{\citenamefont {Grebenev}, \citenamefont {Nazarenko},\ and\
  \citenamefont {Medvedev}(2017)}]{zamm}%
  \BibitemOpen
  \bibfield  {author} {\bibinfo {author} {\bibfnamefont {V.~N.}\ \bibnamefont
  {Grebenev}}, \bibinfo {author} {\bibfnamefont {S.~V.}\ \bibnamefont
  {Nazarenko}}, \ and\ \bibinfo {author} {\bibfnamefont {S.~B.}\ \bibnamefont
  {Medvedev}},\ }\bibfield  {title} {\enquote {\bibinfo {title} {Complementary
  remarks to properties of the energy spectrum in leith's model of
  turbulence},}\ }\href {\doibase https://doi.org/10.1002/zamm.201600060}
  {\bibfield  {journal} {\bibinfo  {journal} {ZAMM - Journal of Applied
  Mathematics and Mechanics / Zeitschrift für Angewandte Mathematik und
  Mechanik}\ }\textbf {\bibinfo {volume} {97}},\ \bibinfo {pages} {664--669}
  (\bibinfo {year} {2017})},\ \Eprint
  {http://arxiv.org/abs/https://onlinelibrary.wiley.com/doi/pdf/10.1002/zamm.201600060}
  {https://onlinelibrary.wiley.com/doi/pdf/10.1002/zamm.201600060} \BibitemShut
  {NoStop}%
\bibitem [{\citenamefont {Clewley}, \citenamefont {LaMar},\ and\ \citenamefont
  {J.}(2007)}]{clewley2007pyds}%
  \BibitemOpen
  \bibfield  {author} {\bibinfo {author} {\bibfnamefont {W.}~\bibnamefont
  {Clewley}, \bibfnamefont {R.and~Sherwood}}, \bibinfo {author} {\bibfnamefont
  {M.}~\bibnamefont {LaMar}}, \ and\ \bibinfo {author} {\bibfnamefont
  {G.}~\bibnamefont {J.}},\ }\href@noop {} {\enquote {\bibinfo {title}
  {{PyDSTool}, a software environment for dynamical systems modeling},}\
  }\bibinfo {howpublished} {\url{http://pydstool.sourceforge.net}} (\bibinfo
  {year} {2007}),\ \bibinfo {note} {[Online; accessed
  18-January-2021]}\BibitemShut {NoStop}%
\bibitem [{\citenamefont {Kuznetsov}(2013)}]{kuznetsov2013elements}%
  \BibitemOpen
  \bibfield  {author} {\bibinfo {author} {\bibfnamefont {Y.}~\bibnamefont
  {Kuznetsov}},\ }\href@noop {} {\emph {\bibinfo {title} {Elements of applied
  bifurcation theory}}},\ Vol.\ \bibinfo {volume} {112}\ (\bibinfo  {publisher}
  {Springer Science \& Business Media},\ \bibinfo {year} {2013})\BibitemShut
  {NoStop}%
\bibitem [{\citenamefont {Connaughton}, \citenamefont {Newell},\ and\
  \citenamefont {Pomeau}(2003)}]{connaughton2003non}%
  \BibitemOpen
  \bibfield  {author} {\bibinfo {author} {\bibfnamefont {C.}~\bibnamefont
  {Connaughton}}, \bibinfo {author} {\bibfnamefont {A.}~\bibnamefont {Newell}},
  \ and\ \bibinfo {author} {\bibfnamefont {Y.}~\bibnamefont {Pomeau}},\
  }\bibfield  {title} {\enquote {\bibinfo {title} {Non-stationary spectra of
  local wave turbulence},}\ }\href@noop {} {\bibfield  {journal} {\bibinfo
  {journal} {Physica D: Nonlinear Phenomena}\ }\textbf {\bibinfo {volume}
  {184}},\ \bibinfo {pages} {64--85} (\bibinfo {year} {2003})}\BibitemShut
  {NoStop}%
\bibitem [{\citenamefont {Balescu}(2003)}]{Balescu1}%
  \BibitemOpen
  \bibfield  {author} {\bibinfo {author} {\bibfnamefont {R.}~\bibnamefont
  {Balescu}},\ }\bibfield  {title} {\enquote {\bibinfo {title} {Drift-wave
  turbulence and zonal flow generation},}\ }\href {\doibase
  10.1103/PhysRevE.68.046409} {\bibfield  {journal} {\bibinfo  {journal} {Phys.
  Rev. E}\ }\textbf {\bibinfo {volume} {68}},\ \bibinfo {pages} {046409}
  (\bibinfo {year} {2003})}\BibitemShut {NoStop}%
\bibitem [{\citenamefont {Balescu}, \citenamefont {Petrisor},\ and\
  \citenamefont {Negrea}(2005)}]{Balescu2}%
  \BibitemOpen
  \bibfield  {author} {\bibinfo {author} {\bibfnamefont {R.}~\bibnamefont
  {Balescu}}, \bibinfo {author} {\bibfnamefont {I.}~\bibnamefont {Petrisor}}, \
  and\ \bibinfo {author} {\bibfnamefont {M.}~\bibnamefont {Negrea}},\
  }\bibfield  {title} {\enquote {\bibinfo {title} {Anisotropic electrostatic
  turbulence and zonal flow generation},}\ }\href {\doibase
  10.1088/0741-3335/47/12/005} {\bibfield  {journal} {\bibinfo  {journal}
  {Plasma Physics and Controlled Fusion}\ }\textbf {\bibinfo {volume} {47}},\
  \bibinfo {pages} {2145--2159} (\bibinfo {year} {2005})}\BibitemShut {NoStop}%
\bibitem [{\citenamefont {Gantmacher}\ and\ \citenamefont
  {Brenner}(2005)}]{gantmacher2005applications}%
  \BibitemOpen
  \bibfield  {author} {\bibinfo {author} {\bibfnamefont {F.~R.}\ \bibnamefont
  {Gantmacher}}\ and\ \bibinfo {author} {\bibfnamefont {J.~L.}\ \bibnamefont
  {Brenner}},\ }\href@noop {} {\emph {\bibinfo {title} {
  Applications of the Theory of Matrices}}}\ (\bibinfo  {publisher} {Courier
  Corporation},\ \bibinfo {year} {2005})\BibitemShut {NoStop}%
\bibitem [{\citenamefont {Hassard}, \citenamefont {Kazarinoff},\ and\
  \citenamefont {Wan}(1981)}]{hassard1981theory}%
  \BibitemOpen
  \bibfield  {author} {\bibinfo {author} {\bibfnamefont {B.~D.}\ \bibnamefont
  {Hassard}}, \bibinfo {author} {\bibfnamefont {N.~D.}\ \bibnamefont
  {Kazarinoff}}, \ and\ \bibinfo {author} {\bibfnamefont {Y.-H.}\ \bibnamefont
  {Wan}},\ }\href@noop {} {\emph {\bibinfo {title} {
  applications of Hopf bifurcation}}},\ Vol.~\bibinfo {volume} {41}\ (\bibinfo
  {publisher} {CUP},\ \bibinfo {year} {1981})\BibitemShut {NoStop}%
\bibitem [{\citenamefont {Holoborodko}(2015)}]{holoborodko2015smooth}%
  \BibitemOpen
  \bibfield  {author} {\bibinfo {author} {\bibfnamefont {P.}~\bibnamefont
  {Holoborodko}},\ }\bibfield  {title} {\enquote {\bibinfo {title} {Smooth
  noise-robust differentiators},}\ }\href
  {http://www.holoborodko.com/pavel/numerical-methods/numerical-derivative/smooth-low-noise-differentiators/}
  {\bibfield  {journal} {\bibinfo  {journal} {accessed 01/19/2021}\ } (\bibinfo
  {year} {2015})}\BibitemShut {NoStop}%
\bibitem [{\citenamefont {Van~Rossum}\ and\ \citenamefont
  {Drake}(2009)}]{python}%
  \BibitemOpen
  \bibfield  {author} {\bibinfo {author} {\bibfnamefont {G.}~\bibnamefont
  {Van~Rossum}}\ and\ \bibinfo {author} {\bibfnamefont {F.}~\bibnamefont
  {Drake}},\ }\href@noop {} {\emph {\bibinfo {title} {Python 3 Reference
  Manual}}}\ (\bibinfo  {publisher} {CreateSpace},\ \bibinfo {address} {Scotts
  Valley, CA},\ \bibinfo {year} {2009})\BibitemShut {NoStop}%
\bibitem [{\citenamefont {Peterson}(2009)}]{peterson2009f2py}%
  \BibitemOpen
  \bibfield  {author} {\bibinfo {author} {\bibfnamefont {P.}~\bibnamefont
  {Peterson}},\ }\bibfield  {title} {\enquote {\bibinfo {title} {{F2PY}: a tool
  for connecting {F}ortran and {P}ython programs},}\ }\href@noop {} {\bibfield
  {journal} {\bibinfo  {journal} {International Journal of Computational
  Science and Engineering}\ }\textbf {\bibinfo {volume} {4}},\ \bibinfo {pages}
  {296--305} (\bibinfo {year} {2009})}\BibitemShut {NoStop}%
\bibitem [{\citenamefont {Simpson}(1988)}]{simpson1988touching}%
  \BibitemOpen
  \bibfield  {author} {\bibinfo {author} {\bibfnamefont {J.}~\bibnamefont
  {Simpson}},\ }\href@noop {} {\emph {\bibinfo {title} {Touching the Void}}}\
  (\bibinfo  {publisher} {Harper Perennial},\ \bibinfo {year}
  {1988})\BibitemShut {NoStop}%
\end{thebibliography}%
	
\appendix
\section{Dynamical system analysis for the second order DAM: digest}
\label{sec:GW_2}

\paragraph{Self-similar solution of the second kind.}
The second-order DAM introduced in section~\ref{sec:Failure}
is the conservation law
\be
	\partial_t N + \partial_\omega Q=0, \text{ with } Q:= - \omega^{3+\rho}N^2 \partial_\omega (\omega^{-\rho} N)
	\label{eq:2ndorder}.
\ee
Pure scaling stationary solutions are either the equilibrium solutions  $N \sim \omega^{\rho}$ or the cascade KZ solution 
$N=(3Q/(3\rho-2))^{1/3} \omega^{-2/3}$. Inverse cascade solutions require  $Q<0$ and, therefore, $\rho<2/3$.

 For $\rho=-1$, Eq.~(\ref{eq:2ndorder}) has an  equilibrium solution with the   scaling $N\sim \omega^{-1}$: This  scaling coincides with that of the direct cascade solution of the fourth-order DAM, for both the 3D GW and the 3D NLS (and in fact all the admissible models with $\gamma =0$).

We look for a self-similar solution of the second-kind  describing finite-time infrared blow-up.
Writing
\be
 	N(t,\omega) = \omega_*^{-x} F(\eta), \text{ with } \eta=\omega/\omega_*, \; \omega_* \sim (t_*-t)^b, \; \text{and  }  b:=\dfrac{1}{2x-1} 
\ee
yields the second-order ODE, in terms of the self-similar profile $F$ and the self-similar flux $Q$
\be
	\begin{cases}
	&\partial_\eta F =-Q \eta^{-3}F^{-2} +\rho \eta^{-1}F\\
	&\partial_\eta Q= b \left( (x-\rho) F + Q \eta^{-2} F^{-2}\right)
	\end{cases},
	\text{with boundary conditions
	$F=Q=0$ at $\eta= 1$ and $\eta \to \infty$}.
\ee
\paragraph{Reduction to an autonomous system.}
The system can be reduced to a second-order autonomous sytem, in terms of the parametrization $\tau=\log \eta$ and the rescaled variables  $f=\eta^{1/2}F$ and $q=\eta^{-1/2}Q$, as
\be
	\begin{cases}
	&f'(\tau) =\left(\rho+\frac{1}{2}\right)f - q f^{-2}, \\
	&	q'(\tau) =-\frac{1}{2}q +b \left(q f^{-2}-(\rho+x)f\right)
	\end{cases}
	\label{eq:autonomous_2}
\ee
with boundary conditions $f=q=0$ at $\tau= 0$ and $\tau\to \infty$.
To remove singularity at the origin, one can rely on the parametrization $\theta :=\int_1^\tau f^{-2} d\tau' $, and obtain the system
\be
	\begin{cases}
	&f'(\theta) =\left(\rho+\frac{1}{2}\right)f^3 -q, \\
	&	q'(\theta) =-\frac{1}{2}qf^2 +b \left(q -(\rho+x)f^3\right)
	\end{cases}
	\label{eq:autonomous_2_nonsing}
\ee

\paragraph{The $\infty_\star$-cycle.}
A unique solution to system (\ref{eq:autonomous_2}) exists for a unique value of the exponent $x < 2/3$, as proven in a more general case \cite{nazarenko2016self}. In previous works \cite{thalabard2015anomalous,galtier2019nonlinear}, the specific  value of $x$ was found using the shooting methods and a different set of rescaled variables  instead of the variables $f$, $q$.
Here, we rely on the  numerical  continuation algorithms provided by the Python  library PyDSTool~\cite{clewley2007pyds} to retrieve those results.
We observe that for $\rho<-1/2$, the system has three fixed points, namely the marginally stable node  $(f,q)=(0,0)$ and the foci $(f_\pm,q_\pm)= \pm (-1/2-\rho)^{-1/2}, -(-1/2-\rho)^{-3/2})$.  

\begin{figure}
	\includegraphics[width=\textwidth,trim=0cm 0cm 0cm 0cm,clip]{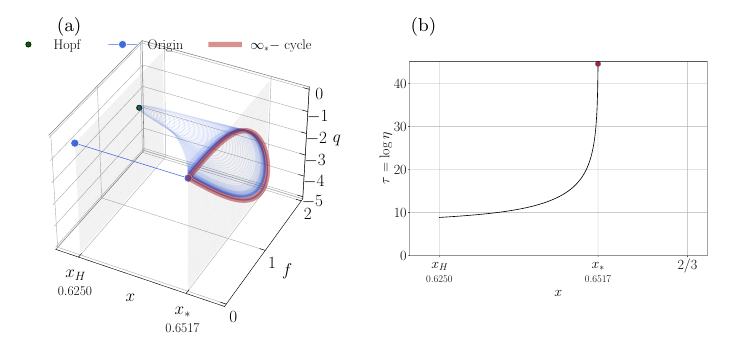}
	\caption{{\bf Numerical continuation for the second-order DAM.} Panel (a):  the periodic orbits of system (\ref{eq:autonomous_2}) growing from the Hopf point at $x=x_H$ to the point $x_*$ where it collides with the marginal node $(0,0)$. The final homoclinic orbit is shown in red. Panel (b) shows  the period of the cycles as a function of $x$.}
	\label{Fig:2d_conti}
\end{figure}

The positive focus $(f_+, q_+)$ undergoes Hopf bifurcation at $x_H \simeq 0.6250$, and gives rise to a branch of stable periodic orbits.
Fig. \ref{Fig:2d_conti} shows the outcome of the continuation algorithm; the left panel  shows that the  branch of periodic solutions stops at $x=x_*\simeq 0.6517$, at which point  the cycle collides with the marginal node $(0,0)$. Besides, the right panel shows that the final cycle has infinite period: $x_*\simeq 0.6517$  is therefore the looked-after exponent, as indeed previously found from the shooting methods~ \cite{galtier2019nonlinear}.
 
 \section{Fixed point analysis and Hopf bifurcation.}
 \label{sec:fixed}
 \label{Unsingular system.}
 The system (\ref{eq:autonomous4}) is singular at the origin $f=g=p=q=0$. To apply the standard fixed-point analysis, we remove the singularity  by changing the time 
 variable $\tau \to \theta=\int_{\tau_0}^\tau f^{-2} (\tau') d \tau'$ (with arbitrary $\tau_0 >0$) thereby obtaining the following system,
\be
 		  \left\{
 			\begin{aligned}
 		 		  f'(\theta) &= f^2 \left(g+ \frac{1}{2}f\right),\\
 		 		  g' (\theta) &= q-p+2fg^2 +\left(\frac{3}{2}-2r\right)gf^2 +r (r-1) f^3, \\
 		 		  p'(\theta) &= -f^2\left(\frac{3}{2}p+b(xf+g)\right),\\
 		 		  q'(\theta) &= -f^2 \left(\frac{1}{2}q+b(xf+g)\right).
 			\end{aligned}
 			 \right.
 			\label{eq:autonomous4_nonsingular}
 \ee
System  	(\ref{eq:autonomous4_nonsingular}) admits a 2D stationary manifold
$
	\mathcal S = \left\{ (f,g,p,q) \in \mathbb R^4: \; 
					f=0, \, p=q
			\right\}
			$
 on which the points have marginal stability. 
Indeed, consider small deviations from a point on the stationary manifold $S$, \emph{i.e.},
 $f=\tilde{f},\quad g=g_0+\tilde{g},\quad p=q_0+\tilde{p},\quad q=q_0+\tilde{q}$.

In terms of the variable $y=2g_0^2f+\tilde{p}-\tilde{q}$, System  (\ref{eq:autonomous4_nonsingular}) becomes
\be
 		  \left\{
 			\begin{split}
 		 		 f^\prime(\theta) &= f^2 \left((g_0+\tilde{g})+ \frac{1}{2}f\right),\\
 		 		 \tilde{g}^\prime(\theta) &=y+2f(\tilde{g}^2+2g_0\tilde{g})+\left(\frac{3}{2}-2r\right)(g_0+\tilde{g})f^2+r(r-1) f^3, \\
 		 		 y^\prime(\theta) &= f^2\left((q_0+\tilde{q})+4g_0^2f+2g_0(g_0+\tilde{g})-\frac{3}{2}y\right),\\
 		 		 \tilde{q}^\prime(\theta) &= -f^2 \left(\frac{1}{2}(q_0+\tilde{q})+b(xf+g_0+\tilde{g})\right).
 			\end{split}
 			 \right.
 			\label{eq:normal_yr}
 		 \ee
\corr Its linearization leads to 
\def\arraystretch{0.1}
\be
\dfrac{d}{d\theta} \tilde Y = L \tilde Y\;\;
\text{with}\;\; \tilde Y = [\tilde f,\tilde g,\tilde y,\tilde q]^T \;\; \text{and}\;\;
L := \begin{bmatrix}
0\;&0\;&0\;&0\\
0\;&0\;&1\;&0\\
0\;&0\;&0\;&0\\
0\;&0\;&0\;&0
\end{bmatrix},
\ee
implying that all the eigenvalues of the linearized system are zero.\rroc~  Therefore, in order to study the dynamics near the stationary manifold $S$, we have to take into account the nonlinear corrections. Only one fixed point on $S$ is relevant to our study, namely the origin
$(f,g,p,q)=(0,0,0,0)$. There are two orbits that enter/leave this point: they correspond to asymptotic behavior of our system at the front $\eta \to 1$ and at \corr the tail \rroc
$\eta \to \infty$. The nonlinear solutions for $\eta \to 1$ and for $\eta \to \infty$ are considered in Appendices~\ref{sec:front} and \ref{sec:etainf} respectively.
In terms of $\theta$, these solutions read
\be
\begin{split}
          (f,g,p,q) \to &
	 \left[ C^{-1/2}\left(-2\theta\right)^{-3/4},  \frac{3}{2}C^{1/2} (-2\theta)^{-1/4},  -\frac{75}{8}C^{3/2} (-2\theta)^{-3/4},
	 -\frac{75}{8}C^{3/2} (-2\theta)^{-3/4} \right]  \\
	 & \text{as} \;\;\; \theta\to -\infty \quad (\eta \to 1).
	\label{eq:fgpq 1 order}
\end{split}
\ee
and
\be
\begin{split}
          (f,g,p,q) \to &
	 \left[  b^{1/2}\theta^{-1/2},\, -xb^{1/2}\theta^{-1/2},\quad  A (3x-2) b^{3/2} \theta^{-3/2},
	 A (3-3x)b^{3/2} \theta^{-3/2}
	 \right]
	 \\
	 & \text{as} \;\;\; \theta\to +\infty \quad (\eta \to \infty)
	\label{eq:fgpq 2 order}
\end{split}
\ee
where $A=(r+x)(r+x-1)$.


When  $|r|>1/2$,  
there exist   two other isolated fixed points, 
\be
	P_\pm= \left( 1,-\dfrac{1}{2},-\dfrac{1}{3},-1 \right) f_\pm \;\; \text{with}\;\; f_\pm := \pm \left(\dfrac{3}{2}\left(r^2-\dfrac{1}{4}\right)\right)^{-1/2},
\label{eq:isolated}
\ee
with only point $P_+$ associated to positive spectrum being physical.
The point $P_+$ is important for our scenario, because it is the change of stability properties of this point that marks the Hopf bifurcation of the limit cycle creation. It is interesting that the position of $P_+$ is independent of the parameter $x$.

The isolated points (\ref{eq:isolated}) exist in the 3D  GW case $(d=3,\alpha=1)$ for which $r=2>1/2$ but not in the  3D NLS case $(d=3,\alpha=2)$ for which $r =1/2$. This means that the fourth-order model for the 3D NLS  is degenerate; \corr one could say that point $P_+$ then formally lies at infinity \rroc.  

Linearizing the system (\ref{eq:autonomous4_nonsingular}) around the stationary point $P_+$, we obtain the following system
\begin{equation}
\tilde{X}^\prime(\theta)=A\tilde{X},    
\end{equation}
where  
$\tilde{X}=\left[\begin{array}{cccc}\tilde{f},&\tilde{g},
&\tilde{p},&\tilde{q},\end{array}\right]^T
$
\corr  denotes the vector \rroc of small perturbations. The matrix $A$ has the following form
\begin{equation}
A=f_+^2\left[\begin{array}{cccc}\frac{1}{2}&1&0&0\\
3r^2-r-1 \;\;&-2r-\frac{1}{2} \;\;  &-f_+^{-2} \;&f_+^{-2}  \\
\frac{x}{1-2x}&\frac{1}{1-2x}&-\frac{3}{2}&0 \\
\frac{x}{1-2x}&\frac{1}{1-2x}&0&-\frac{1}{2}\end{array}\right], 
\end{equation}
Hence, its characteristic polynomial $h(\zeta)$ reads
\begin{equation}\label{A-zE}
h(\zeta)=\det(A-\zeta E)=h_0\zeta^4+h_1\zeta^3+h_2\zeta^2+h_3\zeta+h_4,    
\end{equation}
where
\begin{equation}
h_0=1,\quad h_1=2(r+1)f_+^2,\quad h_2=-2f_+^2+\left(4r+\frac{3}{4}\right)f_+^4,
\end{equation}
\begin{equation}
h_3=(b-4)f_+^{4}+\frac{3}{2}f_+^{6},
\quad h_4=-f_+^6.
\end{equation}
We apply the Routh–-Hurwitz stability criterion~\cite{gantmacher2005applications}:
{\it  All roots of the polynomial $h(\zeta)$ have negative real parts if and only if 
\begin{equation}
\Delta_1=h_1>0,\;\;
\arraycolsep=1.4pt\def\arraystretch{0.1}
\Delta_2=\begin{array}{|cc|}h_1&h_3\\1&h_2\end{array}
\arraycolsep=1.4pt\def\arraystretch{0.1}
>0,\quad
\arraycolsep=1.4pt\def\arraystretch{0.1}
\Delta_3=\begin{array}{|ccc|}h_1&h_3&0\\1&h_2&h_4\\0&h_1&h_3\end{array}>0,
\quad
\Delta_4=h_4\Delta_3>0.
\end{equation}
}

 First,  we consider the case $\Delta_3$ and $\Delta_4 \ne 0$.
Since $h_4<0$, then either $\Delta_3$ or $\Delta_4$ is less than zero. Therefore, the above criterion is not satisfied and all roots of the characteristic polynomial cannot be with negative real parts.
Note that changing the direction of the time-like variable $\theta$ does not change the sign of $h_4=\det(A)=\det(-A)$ and, therefore, does not change the fact of instability. Therefore, we conclude that \corr the neighborhood of $P_+$ \rroc must always include  stable and  unstable orbits simultaneously.

Now consider the case $\Delta_3=0$ (hence $\Delta_4=0$ too). This is, in fact, the case of the Hopf bifurcation $x=x_H$ because it gives the condition which guarantees  existence of  two purely imaginary roots $\zeta_1=i\omega$ and $\zeta_2=-i\omega$ of the characteristic polynomial $h(\zeta)$.
Indeed, it follows from Orlando's formula~ \cite{gantmacher2005applications} for the polynomial $h(\zeta)$ that $\Delta_3=0$ if and only if the sum at least one pair of roots of $h(\zeta)$ is zero. Also since the determinant $h_4=\det(A)$ of the matrix $A$ is nonzero, and recalling that this determinant is equal to the product of all the eigenvalues, we have $\omega \ne 0$. But then, because $h_4=\det(A) <0$, and remembering that the 
roots of the real polynomial $h(\zeta)$ are either real or come in purely imaginary mutually conjugated pairs,  the other two roots $\zeta_3$ and $\zeta_4$ are real numbers of  different signs, $\zeta_3\zeta_4 < 0$.

To find $x_H$, we explicitly compute $\Delta_3$ as
\begin{equation}
f_+^{-6}\Delta_3=-f_+^{-4}b^2-2f_+^{-2}r(3r^2-7r-4)b+36\,r^5-51\,r^4-42\,r^3+\frac{87}{4}\,r^2+\frac {27}{2}r+\frac{3}{4}=0,
\end{equation}
where we recall 
$b=\dfrac{1}{2x-1}$.
For example, for the case of the 3D GW system ($r=2$) we get  $x_H\approx 0.60014$, as identified by the numerical continuation software.

In addition to a pair of purely imaginary eigenvalues, the matrix $A(x_H)$ has an eigenvalue, $\zeta_3$ or $\zeta_4$, with a positive real part. The resulting periodic orbit is unstable, as the eigenvalue with positive real part produces a Floquet exponent which in turn has a positive real part (see Remark 3 on p.20 in book\cite{hassard1981theory}). Changing the direction of time  results in the changing the signs of both $\zeta_3$ and $\zeta_4$. This does not alter the fact that one of these eigenvalues is positive. Thus, the emerging limit cycle is unstable in the case of the reversed time too. This fact makes it practically impossible to find the limit cycle numerically by directly computing the trajectories in the 4D phase space, because only measure-zero set of initial points would lead to orbits attracted to the cycle.
This is another difference with the 2D case \cite{thalabard2015anomalous}.
Therefore, \corr in order to find the cycles, we employ numerical continuation software, and specifically the PyCont library within the  PyDSTool Python environment \cite{clewley2007pyds} which is specially designed for finding unstable limit cycles; see Appendix~\ref{sec:continuation} \rroc.

 \section{Numerics}
 \corr
\subsection{ Numerical simulations and regularization}
\rroc
\label{sec:dnsdetails}
To simulate the transients of the fourth-order DAM,
we use a log-discretization of the frequency spaces, and the grid points $\omega_i =  2^{i/\kappa}, i\in [-1200,1200]$ where the parameter $\kappa=20$ controls the frequency binning. To regularize the system, we do not employ any viscosity but rather rely on  smooth noise-robust differentiators \cite{holoborodko2015smooth} that compute derivatives in frequency space as
\be
	D[ f,i]= \dfrac{42 \delta_1 +48 \delta_2 +23 \delta_3 + 8 \delta_4 + \delta_5}{512 h \omega_i} \;\; \text{with }\;\; \delta_k:= f_{i+k}-f_{i-k} \; \text{and}\; h= \kappa^{-1}\log 2.
\ee
We use the ADAM-Bashforth scheme of second-order to advance in time. Time steps are determined through the  CFL condition:
\be
	\Delta t=  \mu \times \min_\omega \dfrac{ N+\epsilon}{ |D_tN +\epsilon|}
\text{with\; $\epsilon=10^{-40}$ and $\mu =2^{-8}$ (GW) or $2^{-11}$ (NLS)}
\ee
In all the cases reported here, the initial condition for the waveaction density $N(\omega)$ is a Gaussian centered at $\omega=10^{13}$, with amplitude $10^{-5}$ and variance $0.1$. Integration is made with the Python programming language  \cite{python} and sped-up using the \emph{f2py} package\cite{peterson2009f2py}. 

\subsection{Numerical continuation and the Simpson strategy}
\label{sec:continuation}
To ``chase'' the $\infty_\star$-cycle, determine the anomalous exponent and generate the series of Figures~\ref{fig:LimitCycle_GW},
\ref{fig:LimitCycle_NLS}, \ref{fig:LimitCycle_final}, \ref{Fig:2d_conti}, we  relied on the numerical continuation  library PyCont package from the PyDSTool environment \cite{clewley2007pyds}, as described in Appendix~\ref{sec:GW_2}.
Rather than computing its non-singular counterpart, it proved more efficient to deal with  the original singular system~(\ref{eq:autonomous4}).\\

Computing the branches of periodic solutions  is straightforward in the generic case $\alpha \le 1$, which includes the 3D GW case. 
 In that case,  the $\infty_\star$-cycle is grown directly from the Hopf point at fixed $\alpha$. After a series of bifurcations identified by the software as either  \emph{PD} or \emph{LPC} points, the system converges towards $x_\star$, and the convergence is increased upon decreasing the step-sizes and the numerical tolerance of the software. In our interpretation, the PD points flagged by the software appear to be spurious and the LPC points are genuine and represent the fold bifurcations discusses in the main text in relation to Fig.~\ref{fig:LimitCycle_GW}.\\

\begin{figure}
	\includegraphics[width=\columnwidth]{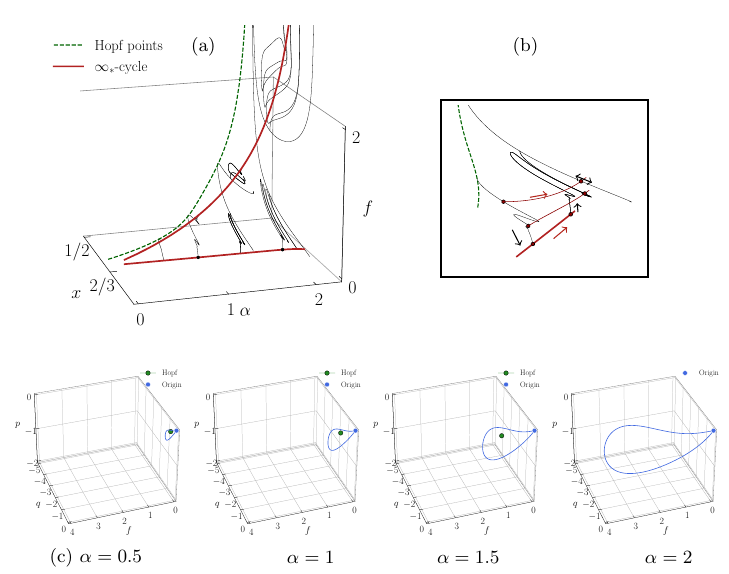}
	\caption{{\bf Numerical continuation of the $\infty_*$-cycle.} Panel (a) shows the extrema of $f$ along the  $\infty_*$-cycle (in red), obtained from the continuation of the GW $\infty_*$-cycle by continuous alteration of $\alpha $ and $x_*$. The black lines are continuation at fixed value of $\alpha$, as in Fig.~\ref{fig:LimitCycle_GW} and 	Fig.~\ref{fig:LimitCycle_NLS}. Panel (b) illustrates the Simpson strategy described in \S \ref{sec:continuation} to find the branches of periodic solutions for $\alpha >1$. Panel (c) shows the  $\infty_*$-cycles for various values of $\alpha$.}
	\label{fig:LimitCycle_bridge}
\end{figure}

Computing the branches of periodic solutions  is less straightforward when  $\alpha > 1$ which, in particular, includes the 3D NLS case.
\corr Note that this range includes the non-generic cases $\alpha \ge 2$ but also the generic cases $1<\alpha<2\rroc$ for which the Hopf point is well-defined. \rroc
Numerical continuations starting from the Hopf point then either fail to converge towards the $\infty_\star$-cycle or simply cannot be initiated due to the Hopf points lying at $\infty$.  To generate the rather intricate patterns of Fig.~\ref{fig:LimitCycle_NLS}, the \emph{Simpson strategy} \cite{simpson1988touching}, however, proves very fruitful. The \emph{Simpson strategy} consists of the three following steps illustrated in Panel (b)	of Fig.~ \ref{fig:LimitCycle_bridge} \emph{(i)} Grow the cycles  at fixed $\alpha=1$ until the $\infty_\star$ -cycle is converged  (up to some thresholds prescribed by  the tolerance parameters), \emph{(ii)} Bridge the 3D GW $\infty_\star$-cycle to the  desired $\alpha>1$ by performing continuation with both $\alpha$ and $x$ as free parameters, \emph{(iii)} at desired $\alpha$, freeze $\alpha$ and either grow or shrink the resulting cycle to check convergence towards $\infty_\star$-cycle and generate associated branch of non-infinite periodic solutions.  The outcome of the Simpson strategy is represented by the red lines in Panel~ (a) of Fig.\ref{fig:LimitCycle_bridge} which determine the behavior of $ x_*$ in the final phase portrait of Fig.~\ref{fig:LimitCycle_final}.\\

To generate the full patterns of periodic branches at fixed $\alpha>1$ as shown {\em e.g.} in Fig.~\ref{fig:LimitCycle_NLS} and in Panel~(a) of   Fig.~\ref{fig:LimitCycle_bridge}, the same strategy is performed by modifying the step  \emph{(i)} of the Simpson strategy to stop the growth of the 3D GW cycle to some determined finite sizes rather than  to the $\infty_\star$ cycle.
All the numerics were done using a standard Dell XPS 13 laptop computer.

\section{Asymptotics near the origin, $(f,g,p,q) \to 0$}

Two relevant asymptotics near the fixed point $(f,g,p,q) \to 0$ correspond to the 
sharp  front at $\eta \to 1$ and the power-law asymptotics and the power-law asymptotic at
large $\eta$. Below, it will be easier for us to work with the original profiles 
$F, G=F', P, Q$ and the similarity variable $\eta$.

\subsection{Solution near the sharp front, $\eta \to 1$.}
\label{sec:front}
Equation (\ref{eq:selfsimilarF}) can be represented as a system of four first-order differential equations for the  variables $F$, $G=F'$, $P$ and $Q$:
\begin{equation}\label{FPQsystem}
F'=G,\quad Q'=-b\,(xF+\eta G),\quad P'=-\eta b\,(xF+\eta G),
\end{equation}
\begin{equation}\label{eqG}
\eta^5 F^2G\,'=r(r-1)\eta^3F^3+2\eta^4FG(\eta G-rF)+\eta Q-P.
\end{equation}

We seek a solution near the frontal point, $\eta \to 1$ in the following asymptotic form
\begin{equation}\label{expansionF0}
F(\eta)=C(\eta-1)^{\mu}
\end{equation}
with the positive constants $C$ and $\mu$ to be found.
Substituting this representation into the first equation in (\ref{FPQsystem}), we have
\begin{equation}\label{F0G}
G(\eta)=C\mu(\eta-1)^{\mu-1}.
\end{equation}
Substituting (\ref{expansionF0}) and (\ref{F0G}) into the two equations for $P$ and $Q$, and integrating them, we get
\begin{equation}
P=C_P-C\,b (\eta-1)^{\mu}-\frac{2\mu+x}{\mu+1}C\,b\,(\eta-1)^{\mu+1}-
\frac{\mu+x}{\mu+2}C\,b\,(\eta-1)^{\mu+2},
\end{equation}
\begin{equation}
Q=C_Q-C\,b(\eta-1)^{\mu}-C\,b\frac{\mu+x}{\mu+1}(\eta-1)^{\mu+1},
\end{equation}
where $C_P$ and $C_Q$ are constants. 
We are looking for a solution with zero fluxes $P$ and $Q$ at $\eta = 1$, so we put $C_P=C_Q=0$. Substituting the expressions for $F$, $G$, $P$ and $Q$ in (\ref{eqG}), we get in the leading order in $(\eta -1)$:
\begin{equation}
\frac{C\,b}{\mu+1}(\eta-1)^{\mu+1}=C^3\mu(\mu+1)(\eta-1)^{3\mu-2}.
\end{equation}
To satisfy this equation, we must choose
\begin{equation}
\mu=\frac{3}{2},\quad C=\sqrt{\frac{8\,b}{75}}.
\end{equation}

\subsection{Power-law asymptotics for $\eta \to\infty$.}
\label{sec:etainf}

Consider  the solution for large $\eta$
in a power law form,
\begin{equation}
F=C\eta^{-\nu},\quad C>0,\quad \nu>0.
\end{equation}
We proceed in the same way as for the frontal point. We substitute $F(\eta)$ into the first equation in (\ref{FPQsystem}) and find {the function} $G(\eta)$. Then we substitute {the functions } $F(\eta)$ and $G(\eta)$ into the second and third equations in (\ref{FPQsystem}) and, integrating them, find $Q$ and $P$. Substituting the calculated functions in (\ref{eqG}), we get
\begin{equation}
C^3(\nu+r)(\nu+r-1)\eta^{3-3\nu}+Cb\frac{\nu-x}{(\nu-1)(\nu-2)}\eta^{2-\nu}+C_Q\eta-C_P=0,
\end{equation}
where $C_Q$ and $C_P$ are the constants.

The first term { has the leading order} for $\nu\in(0,1/2)$. But the prefactor for this term can be zero only for $\nu=-r$ or $\nu=1-r$.
The second term is the leading one for $\nu\in(1/2,1)$. To nullify this term, we must have $\nu=x$. The third term is { of the leading order} for $\nu>1$. To cancel this term   we must set $C_Q=0$.

The two boundary values $\nu=1/2$ and $\nu=1$ require separate consideration. We have $C=4/\sqrt{24r^2-6}$ for $\nu=1/2$. For $\nu=1$, the leading term vanishes only for $x=1$.
Thus, our choice of the  large $\eta$ asymptotics of $F(\eta)$ as a power law with the exponent $\nu = x$ is consistent with the values of $x$ in the interval $(1/2,1)$.

\end{document}